\documentclass[11pt, a4paper]{article}
\pdfoutput=1

\usepackage[height=8.85in,width=6.55in]{geometry}

\usepackage{graphicx,rotating}     
\usepackage[bookmarksopen,colorlinks=true,linkcolor=steelblue,
citecolor=darkred,urlcolor=darkred,linktoc=all]{hyperref}

\usepackage{amsmath,amssymb,subdepth}
\usepackage{mathtools}
\usepackage{mathrsfs}
\usepackage{slashed}
\usepackage{bm}
\usepackage{cite}
\usepackage{comment}
\usepackage{cleveref}
\usepackage[footnotesize]{caption}
\usepackage{cancel}
\usepackage{here}
\usepackage{booktabs}
\usepackage{subcaption}

\usepackage[T1]{fontenc}
\usepackage[utf8]{inputenc}

\usepackage{trimclip}
\usepackage[framemethod=default]{mdframed}
\newmdenv[skipabove=7pt,
skipbelow=7pt,
rightline=false,
leftline=false,
topline=false,
bottomline=false,
backgroundcolor=gray!10,
linecolor=gray,
innerleftmargin=5pt,
innerrightmargin=5pt,
innertopmargin=5pt,
innerbottommargin=5pt,
leftmargin=0cm,
rightmargin=0cm,
linewidth=4pt]{eBox}
\newmdenv[skipabove=7pt,
skipbelow=7pt,
rightline=true,
leftline=true,
topline=true,
bottomline=true,
backgroundcolor=white,
linecolor=gray,
innerleftmargin=5pt,
innerrightmargin=5pt,
innertopmargin=5pt,
innerbottommargin=5pt,
leftmargin=0cm,
rightmargin=0cm,
linewidth=1pt]{eBox2}

\usepackage{chngcntr}
\counterwithin{equation}{section}

\usepackage{xcolor}
\definecolor{darkred}{rgb}{0.7, 0., 0.}
\definecolor{orangered}{rgb}{1,0.27,0.}
\definecolor{steelblue}{rgb}{0.275,0.51, 0.706}
\definecolor{forestgreen}{rgb}{0.13,0.55,0.13}
\definecolor{violet}{cmyk}{.32,.95,.17,.00}

\definecolor{sagegreen}{rgb}{0.5, 0.65, 0.4}
\definecolor{sepia}{rgb}{0.55, 0.45, 0.3}
\newcommand{\mpl}{M_{\mathrm{Pl}}}
\newcommand{\TR}{\mathrm{T}_{\rm R}}
\newcommand{\mphi}{m_\phi}
\newcommand{\Gphi}{\Gamma_\phi}
\newcommand{\ii}{\mathrm{i}}

\newcommand{\dA}{d_{\rm A}}

\newcommand{\dd}{\mathop{}\!\mathrm{d}}
\newif\ifkeepBg
\keepBgfalse
\newcommand{\Bgfac}{\ifkeepBg B_g\,\fi}

\begin{document}
\hypersetup{pageanchor=false}
\begin{titlepage}

\begin{center}

\hfill KEK-TH-2868

\vskip 1in

{\Huge \bfseries
Gravitational Waves from Reheating\\
beyond Instantaneous Thermalization\\
}
\vskip .8in

{\Large
Kyohei Mukaida$^{\blacklozenge\lozenge}$,
Tenta Tsuji$^{\blacklozenge\lozenge*}$
}

\vskip .3in

\begin{tabular}{@{}c@{\hspace{0.65em}}p{0.90\textwidth}@{}}

$\blacklozenge$
&
\emph{Theory Center, IPNS, KEK,
1-1 Oho, Tsukuba, Ibaraki 305-0801, Japan}
\\[0.4em]

$\lozenge$
&
\emph{The Graduate University for Advanced Studies (SOKENDAI),
1-1 Oho, Tsukuba, Ibaraki 305-0801, Japan}

\end{tabular}

\end{center}
\vskip .5in

\begin{abstract}
\noindent
We study gravitational-wave production during perturbative reheating without
assuming instantaneous thermalization of the inflaton decay products.
The injected energetic particles thermalize through an in-medium cascade of nearly collinear splittings subject to the Landau--Pomeranchuk--Migdal effect and are ultimately isotropized by elastic scatterings near the thermal scale.
The non-thermal hard population present before complete thermalization produces an additional
gravitational-wave component through hard--soft scatterings, with
$\Omega_{\rm GW}\propto f^{1/2}$ below the injection-scale turnover.
The same isotropization erases the directional information underlying the
vacuum $1/k$ bremsstrahlung soft pole, changing the deep-infrared spectrum
from $\Omega_{\rm GW}\propto f$ to $\Omega_{\rm GW}\propto f^3$.
\end{abstract}

\vfill

\noindent
{\footnotesize
$^{*}$\,\textit{E-mail:}\,
\href{mailto:tenta@post.kek.jp}{\nolinkurl{tenta@post.kek.jp}}
}

\end{titlepage}

\tableofcontents
\renewcommand{\thepage}{\arabic{page}}
\renewcommand{\thefootnote}{$\natural$\arabic{footnote}}
\setcounter{footnote}{0}
\hypersetup{pageanchor=true}

\newpage
\section{Introduction}
\label{sec:introduction}

Inflation provides a compelling framework for the origin of the hot Big Bang, resolving the horizon and flatness problems while generating the primordial perturbations that seed cosmic structure.  After inflation, the energy stored in the inflaton must be transferred to relativistic particles and ultimately converted into a thermal plasma through reheating~\cite{Allahverdi:2010xz,Amin:2014eta}.  Successful Big Bang nucleosynthesis and related cosmological constraints
require reheating to have produced a sufficiently thermalized radiation bath
by temperatures of order a few MeV
~\cite{
KawasakiKohriSugiyama:1999,
Kawasaki:2000en,
Hannestad:2004Reheating,
deSalas:2015glj,
Hasegawa:2019jsa,
Barbieri:2025moq
},
but provide little information about how reheating and thermalization proceeded
at earlier times.  The microscopic history connecting the end of inflation to the onset of the
radiation-dominated hot Big Bang remains largely unconstrained observationally
~\cite{Amin:2014eta,Allahverdi:2020bys}.

Gravitational waves provide one of the few direct messengers of this otherwise inaccessible epoch. Once produced, gravitons interact only gravitationally and free-stream to the present Universe, retaining spectral information about the physical scales and expansion history at their production. This has motivated growing interest in high-frequency gravitational waves from reheating. Two generic contributions have been studied extensively: graviton production from the thermal plasma~\cite{Ghiglieri:2015nfa,Ghiglieri:2020mhm,Castells-Tiestos:2022qgu,Ghiglieri:2022rfp,Muia:2023wru,Drewes:2023oxg,Ghiglieri:2024ghm,Ringwald:2020ist,Klose:2022knn,Bernal:2024jim}, and graviton bremsstrahlung associated with energetic inflaton decays~\cite{Nakayama:2018ptw,Huang:2019lgd,Barman:2023ymn,Barman:2023rpg,Jiang:2024akb,Bernal:2023wus,Tokareva:2023mrt,Xu:2024fjl,Xu:2025wjq,Cline:2026jra}. Both arise from the universal coupling of gravity to the energy--momentum tensor and provide complementary probes of the reheating plasma and its microscopic production history\footnote{Gravitons are also produced by the gravitational annihilation of the coherently oscillating inflaton condensate, $\phi\phi\to hh$~\cite{Ema:2020ggo,Xu:2024fjl}. This contribution is typically subdominant to graviton bremsstrahlung in the regime considered here.}.

For graviton bremsstrahlung from inflaton decay, the infrared behavior is
controlled by the gravitational soft theorem~\cite{Weinberg:1965nx}.
In the soft limit, the emission amplitude from energetic decay products exhibits
the familiar $1/k$ soft pole, where $k$ is the graviton momentum.
Existing calculations of inflaton-decay bremsstrahlung are formulated in vacuum,
with the daughter particles treated as freely propagating asymptotic
states~\cite{Nakayama:2018ptw,Huang:2019lgd,Barman:2023ymn,
Barman:2023rpg,Jiang:2024akb,Bernal:2023wus,Tokareva:2023mrt,Xu:2024fjl,Xu:2025wjq,Cline:2026jra}.
In this setup, the soft pole can be understood as a consequence of the
persistent directional flow of energy carried by the outgoing particles.
The decay products that populate the Standard-Model plasma do not
propagate freely.

The thermalization of energetic particles coupled to the Standard-Model plasma
proceeds through an in-medium cascade.  For momenta well above the plasma
temperature, energy loss is dominated by successive nearly collinear
splittings subject to the Landau--Pomeranchuk--Migdal (LPM) effect, which
transport the injected energy toward lower momenta
~\cite{Harigaya:2013vwa,Mukaida:2015ria,Mukaida:2022bbo,Mukaida:2024jiz}.
As these splittings are nearly collinear, the cascade retains the
directional flow of the original energetic particles throughout the hard
regime.  Once the cascade reaches momenta of order the plasma temperature, elastic
scatterings efficiently randomize the particle directions and erase this
directional information~\cite{Mehtar-Tani:2022zwf,Schlichting:2020lef}.
What, then, becomes of the soft gravitational-bremsstrahlung spectrum once this finite lifetime of directional information is taken into account?

Finite-time thermalization has another consequence.  Since inflaton decay
continuously injects energetic particles while their cascade toward the
thermal scale takes a finite time, a dilute non-thermal hard population
coexists with the thermal bath throughout reheating
~\cite{Harigaya:2013vwa,Mukaida:2015ria,Mukaida:2022bbo,Mukaida:2024jiz}.
Such hard particles are known to have important consequences for the production
of weakly coupled species, including dark matter, because they probe
center-of-mass energies far above those available in the thermal bath~\cite{Harigaya:2019tzu}.
The same observation applies to gravitons: scatterings between the hard
particles and the thermal plasma provide an additional source of
gravitational waves that is absent in the instantaneous-thermalization
description.

In this paper, we study gravitational-wave production during perturbative
reheating without assuming instantaneous thermalization.  We consider two
consequences of the finite-time thermalization of energetic inflaton-decay
products.  First, the non-thermal hard population sustained during the
thermalization cascade provides an additional source of gravitons through
scatterings with the thermal bath.  We find that this contribution produces a
high-frequency non-thermal component with
$\Omega_{\rm GW}\propto f^{1/2}$ below the injection-scale turnover.
Second, we revisit graviton bremsstrahlung from inflaton decay, taking
into account the finite lifetime of the directional information carried by the
decay products.  This removes the vacuum soft behavior in the deep infrared,
changing the spectrum from $\Omega_{\rm GW}\propto f$ to
$\Omega_{\rm GW}\propto f^3$ below a thermalization-controlled turnover.

The paper is organized as follows.
In Sec.~\ref{sec:thermalization}, we review perturbative reheating and the
thermalization of energetic decay products through the LPM cascade.
In Sec.~\ref{sec:thermal-baseline}, we summarize the gravitational-wave
spectrum in the instantaneous-thermalization approximation, which serves as
our baseline.
In Sec.~\ref{sec:beyond}, we compute the additional gravitational-wave
production from the non-thermal hard population.
In Sec.~\ref{sec:bremss-finite-thermalization}, we study how finite-time thermalization modifies graviton bremsstrahlung from inflaton decay. In Sec.~\ref{sec:Results}, we summarize numerical results. We conclude in Sec.~\ref{sec:conclusion}. 
Technical details are collected in the appendices.

\section{Perturbative reheating and thermalization}
\label{sec:thermalization}
\subsection{Quadratic, constant-width reheating}
\label{sec:background}

We first specify the reheating background used throughout this paper. We assume that, over the field range relevant for reheating, the inflaton potential is well
approximated by
\begin{equation}
 V(\phi)\simeq \frac{1}{2}m_\phi^2\phi^2.
\end{equation}
Once the oscillation-averaged description is applicable,
$m_\phi\gg H,\Gamma_\phi$, the homogeneous inflaton background has the averaged
equation-of-state parameter $\langle w_\phi\rangle\simeq0$ and behaves as
pressureless matter
~\cite{Turner:1983he,Shtanov:1994ce,Amin:2014eta}.
We restrict attention to the perturbative regime in which nonperturbative
particle production and medium corrections to the decay rate can be neglected,
and take the perturbative decay width $\Gamma_\phi$ to be constant
~\cite{Kofman:1994rk,Kofman:1997yn,Shtanov:1994ce,Amin:2014eta,
Mukaida:2012qn,Mukaida:2012bz,Drewes:2013iaa}.
This regime is naturally realized for sufficiently weak inflaton couplings,
for example through dimension-five interactions suppressed by the Planck scale.
Such weak couplings suppress nonperturbative particle production, while
$\Gamma_\phi\sim m_\phi^3/\mpl^2$ implies a reheating scale well below
$m_\phi$, making thermal corrections to the inflaton dissipation rate
typically small.

For a constant decay width $\Gphi$, the evolution of the homogeneous inflaton background and radiation is
described by the Boltzmann equations~\cite{Albrecht:1982mp,Chung:1998rq,Giudice:2000ex,Allahverdi:2010xz,Garcia:2020eof,Garcia:2020wiy}:
\begin{align}
 \dot\rho_\phi+3H\rho_\phi&=-\Gphi\rho_\phi,
 \label{eq:rho-phi}\\
 \dot\rho_R+4H\rho_R&=\Gphi\rho_\phi,
 \label{eq:rho-r}
\end{align}
where the Hubble parameter is determined by the Friedmann equation
\begin{equation}
H^2=\frac{\rho_\phi+\rho_R}{3\mpl^2}.
\label{eq:friedmann}
\end{equation}
Here $\rho_R$ denotes the total energy density carried by relativistic
decay products, independently of whether they have thermalized, and
$\mpl=(8\pi G_N)^{-1/2}$ is the reduced Planck mass.

The conventional reheating-temperature parameter is defined by
\begin{equation}
 \TR\equiv
 \left(\frac{90}{\pi^2g_{*,R}}\right)^{1/4}
 \sqrt{\Gphi\mpl}.
 \label{eq:TR-convention}
\end{equation}
This conventional definition is widely used in perturbative-reheating
analyses~\cite{Chung:1998rq,Giudice:2000ex,Allahverdi:2010xz,
Harigaya:2013vwa,Garcia:2020eof,Garcia:2020wiy}.
Before equilibration, $\TR$ is a parameter fixed by $\Gphi$, rather than
the physical temperature of the soft sector.  For the Planck-suppressed
regime discussed above, Eq.~\eqref{eq:TR-convention} gives parametrically
$\TR/m_\phi\sim(m_\phi/\mpl)^{1/2}\ll1$.
Planck-suppressed reheating naturally realizes a large hierarchy between
the injection scale $p_0\simeq m_\phi/2$ and the conventional reheating
scale $\TR$.

Deep in the matter-dominated era and after the initial temperature rise, one recovers the familiar approximate scalings
\begin{equation}
 H\propto a^{-3/2},
 \qquad
 \rho_R\propto a^{-3/2},
 \qquad
 T_{\rm eq}\equiv
 \left(\frac{30\rho_R}{\pi^2g_*}\right)^{1/4}
 \propto a^{-3/8}.
 \label{eq:standard-reheating-scaling}
\end{equation}
These matter-dominated reheating scalings are the standard constant-width result~\cite{Chung:1998rq,Giudice:2000ex,Allahverdi:2010xz,Co:2020xaf,Garcia:2020eof,Garcia:2020wiy}.
Here $T_{\rm eq}$ is the temperature that the total radiation energy density would correspond to in equilibrium, and need not coincide with the actual soft-sector temperature before thermalization~\cite{Davidson:2000er,Allahverdi:2002pu,Harigaya:2013vwa,Mukaida:2015ria,Mukaida:2024jiz}.

\subsection{Thermalization during and after reheating}
\label{sec:cascade}

We describe the thermalization of the particles produced during
perturbative reheating in kinetic theory.  For simplicity, we take the
perturbative reheating channel to be $\phi\to gg$, with
$\Gamma_{\phi\to gg}=\Gamma_\phi$, so that each decay injects a back-to-back
gluon pair with momentum $p_0\simeq m_\phi/2$.
For a general decay pattern, the source must be decomposed into the primary
species and the corresponding in-medium cascades evolved separately.
The subsequent discussion also applies qualitatively to other
Standard-Model decay products~\cite{Drees:2022vvn,Mukaida:2022bbo}.
We denote by $f_g(t,\bm p)$ the homogeneous ensemble-averaged gluon
distribution.  Since the inflaton background is homogeneous and the decay
axes are randomly oriented, $f_g$ is isotropic on the ensemble level, even
though each individual inflaton decay initially produces a back-to-back pair
with a definite direction.

The evolution of $f_g$ in the expanding background is described by
\begin{equation}
 \left(\partial_t-Hp\partial_p\right)f_g(t,\bm p)
 =
 C[f_g](t,\bm p)+S_\phi(t,\bm p),
 \label{eq:gluon-kinetic}
\end{equation}
where $C[f_g]$ denotes the collision term generated by interactions among the
plasma constituents, and $S_\phi(t,\bm p)$ denotes the phase-space source from
inflaton decay.

To make the structure of the source explicit, consider first a single
inflaton decay with axis $\hat{\bm n}$.  We define its momentum-space
injection profile by
\begin{equation}
 \mathcal I_{\hat{\bm n}}(\bm p)
 \equiv
 \frac{(2\pi)^3}{\nu_g}
 \left[
  \delta^{(3)}(\bm p-p_0\hat{\bm n})
  +
  \delta^{(3)}(\bm p+p_0\hat{\bm n})
 \right],
 \label{eq:single-decay-injection}
\end{equation}
where $\nu_g$ is the number of gluon color and helicity states.  With
$\int_{\bm p}\equiv\int\dd^3p/(2\pi)^3$, this profile is normalized as
\begin{equation}
 \nu_g\int_{\bm p}\mathcal I_{\hat{\bm n}}(\bm p)=2,
 \qquad
 \nu_g\int_{\bm p}p\,\mathcal I_{\hat{\bm n}}(\bm p)
 =2p_0\simeq m_\phi .
 \label{eq:single-decay-injection-normalization}
\end{equation}
\ifkeepBg
Defining the gluonic branching fraction by
$B_g\equiv\Gamma_{\phi\to gg}/\Gamma_\phi$,
\else
With $\Gamma_{\phi\to gg}=\Gamma_\phi$ in our working model,
\fi
the ensemble-averaged source is
\begin{align}
 S_\phi(t,\bm p)
 &=
 \Bgfac\Gamma_\phi\frac{\rho_\phi(t)}{m_\phi}
 \int\frac{\dd\Omega_{\hat{\bm n}}}{4\pi}\,
 \mathcal I_{\hat{\bm n}}(\bm p)
 \nonumber\\
 &=
 \frac{4\pi^2\Bgfac\Gamma_\phi\rho_\phi(t)}
 {\nu_g m_\phi p_0^2}\,
 \delta(p-p_0).
 \label{eq:inflaton-source-decomposition}
\end{align}
The orientation average makes $S_\phi$ isotropic, although each individual
decay injects a back-to-back pair along a definite axis.  We return to the
unaveraged injection $\mathcal I_{\hat{\bm n}}$ in
Sec.~\ref{sec:bremss-finite-thermalization}.

At weak coupling, the collision term is governed by two classes of processes,
\begin{equation}
 C[f_g]
 =
 C_{2\leftrightarrow2}[f_g]
 +
 C_{1\leftrightarrow2}[f_g].
 \label{eq:collision-term-decomposition}
\end{equation}
Before discussing the thermalization history during reheating, we first
summarize the physical roles of these two terms.
Elastic $2\leftrightarrow2$ scatterings exchange momentum among particles and
relax their angular distribution, whereas nearly collinear
$1\leftrightarrow2$ processes transfer energy from hard particles toward
lower momenta.
We discuss these two ingredients in turn and then return to their combined
effect on the plasma produced by inflaton decay
~\cite{Arnold:2002zm,Harigaya:2013vwa,Mukaida:2015ria,Mukaida:2024jiz}.

We first consider the elastic dynamics described by
$C_{2\leftrightarrow2}$.  Gauge-mediated elastic scattering is enhanced by
screened $t$-channel exchange and is dominated by momentum transfers
$q\ll p$.
For the purpose of describing momentum broadening, the action of
$C_{2\leftrightarrow2}$ on a gluonic excitation propagating through the
ambient medium can be written schematically in the gain--loss form
\begin{equation}
 C_{2\leftrightarrow2}^{\rm el}[f](\bm p)
 \simeq
 \int \dd^3q\,
 \left[
  \frac{\dd\Gamma_{\rm el}(\bm p+\bm q,\bm q)}{\dd^3q}
  f(\bm p+\bm q)
  -
  \frac{\dd\Gamma_{\rm el}(\bm p,\bm q)}{\dd^3q}
  f(\bm p)
 \right].
 \label{eq:elastic-gain-loss}
\end{equation}
Here $\bm q$ is the momentum transferred from the energetic particle to
the medium.

Since the dominant transfers are soft, the gain term in
Eq.~\eqref{eq:elastic-gain-loss} can be expanded in $\bm q$.  With the
quantities inside square brackets evaluated at $(\bm p,\bm q)$,
\[
\begin{aligned}
&
\frac{\dd\Gamma_{\rm el}(\bm p+\bm q,\bm q)}{\dd^3q}
f(\bm p+\bm q)
\\
&\quad =
\frac{\dd\Gamma_{\rm el}(\bm p,\bm q)}{\dd^3q}f(\bm p)
+q_i\partial_{p_i}
\left[
\frac{\dd\Gamma_{\rm el}}{\dd^3q}f
\right]
+\frac12q_iq_j\partial_{p_i}\partial_{p_j}
\left[
\frac{\dd\Gamma_{\rm el}}{\dd^3q}f
\right]
+\cdots .
\end{aligned}
\]
The zeroth-order term cancels the loss term in
Eq.~\eqref{eq:elastic-gain-loss}.  The first moment in $\bm q$ gives momentum
drag, while the second gives momentum diffusion.  This expansion is the
small-angle leading-log Fokker--Planck limit of the elastic collision term.

The transverse second moment of the elastic rate defines the
momentum-broadening coefficient,
\begin{equation}
 \hat q
 \equiv
 \int \dd^2q_\perp\,
 q_\perp^2
 \frac{\dd\Gamma_{\rm el}}{\dd^2q_\perp},
 \qquad
 \left\langle\Delta p_\perp^2(t)\right\rangle
 \simeq
 \hat q\,t .
 \label{eq:qhat-elastic-definition}
\end{equation}
Repeated elastic scatterings randomize the momentum direction.  Since
$\Delta\theta^2\simeq\Delta p_\perp^2/p^2$, their characteristic angular
relaxation rate scales as
\begin{equation}
 \Gamma_{\rm ang}(p)
 \sim
 \frac{\hat q}{p^2}.
 \label{eq:soft-angular-parametric-rate}
\end{equation}

Once the low-momentum sector is approximately thermal, we characterize it
by a temperature $T_s(t)$.  For a weakly coupled thermal bath,
\begin{equation}
 \hat q(T_s)
 \sim
 D_s\alpha_s^2T_s^3,
 \qquad
 D_s\sim C_A^2
 \quad\text{for pure Yang--Mills},
 \label{eq:qhat-thermal-estimate}
\end{equation}
up to logarithmic and order-one corrections
~\cite{Arnold:2002zm,Harigaya:2013vwa,Mukaida:2015ria}.
Substitution into Eq.~\eqref{eq:soft-angular-parametric-rate} gives
\begin{equation}
 \Gamma_{\rm ang}(p)
 \sim D_s\alpha_s^2 T_s
 \left(\frac{T_s}{p}\right)^2 .
\end{equation}
At $p\sim T_s$, angular relaxation occurs on the time scale
$\Gamma_{\rm ang}^{-1}(T_s)\sim(D_s\alpha_s^2T_s)^{-1}$.  When
$\Gamma_{\rm ang}(T_s)\gg H$, elastic scattering establishes kinetic and
angular equilibrium in the low-momentum sector.  Together with efficient
number-changing processes, this allows the soft population to approach local
thermal equilibrium.

We decompose the ensemble-averaged gluon distribution into soft and hard
components,
\begin{equation}
 f_g(t,p)=f_s(t,p)+f_h(t,p),
 \qquad
 f_s(t,p)=n_B\!\left(\frac{p}{T_s(t)}\right),
 \qquad
 n_B(x)\equiv\frac{1}{e^x-1},
 \label{eq:soft-hard-decomposition}
\end{equation}
where $f_s$ denotes the approximately thermal soft component and $f_h$ the remaining non-thermal hard component.
The formation of the soft bath does not imply that the full distribution has
thermalized.  For $p\gg T_s$, the angular-relaxation rate
$\Gamma_{\rm ang}(p)\propto p^{-2}$ becomes progressively smaller, and an
individual small-angle elastic collision transfers only a small fraction of a
hard particle's energy.  Medium-induced inelastic splittings instead control
the degradation of the hard population.

The formation physics of the inelastic $C_{1\leftrightarrow2}$ term can be
seen by considering an energetic parent gluon with momentum $p$ that emits a
daughter gluon with energy $\omega<p$ and transverse momentum $k_\perp$
relative to the parent direction.
The emission is resolved only after the relative phase between the parent and
daughter becomes of order unity.  For a relativistic daughter emitted at a
small angle $\theta\simeq k_\perp/\omega$, the accumulated phase over a time
interval $t$ scales as
\begin{equation}
 \Delta\varphi
 \sim t\,\omega\theta^2
 \sim t\,\frac{k_\perp^2}{\omega}.
\end{equation}
The splitting becomes resolved when $\Delta\varphi \gtrsim 1$.
The corresponding formation time is
\begin{equation}
 t_{\rm form}(\omega)\sim\frac{\omega}{k_\perp^2}.
 \label{eq:lpm-formation-time}
\end{equation}
When several soft scatterings occur within this formation time, their emission
amplitudes interfere coherently, producing the LPM effect
~\cite{Migdal:1956tc,Baier:1996kr,Baier:1996sk,Zakharov:1996fv,
Arnold:2001ba,Arnold:2002ja,Arnold:2008zu,Mukaida:2024jiz}.

During the formation time, repeated soft scatterings broaden the transverse
momentum of the emitted gluon according to
\begin{equation}
 k_\perp^2(t_{\rm form})\sim \hat q\,t_{\rm form}.
\end{equation}
Combining this relation with
$t_{\rm form}\sim\omega/k_\perp^2$ gives
\begin{equation}
 t_{\rm form}(\omega)\sim
 \sqrt{\frac{\omega}{\hat q}}.
\end{equation}
A resolved splitting occurs with probability of order $C_A\alpha_s$ per
formation time, where $\alpha_s$ is the strong coupling and $C_A$ is the
adjoint quadratic Casimir.  The corresponding splitting rate at daughter
energy $\omega$ is
\begin{equation}
 \Gamma_{\rm split}(\omega)
 \sim
 \frac{C_A\alpha_s}{t_{\rm form}(\omega)}
 \sim
 C_A\alpha_s\sqrt{\frac{\hat q}{\omega}},
\end{equation}
up to logarithmic and order-one corrections
~\cite{Baier:1996kr,Zakharov:1996fv,Arnold:2002ja,Arnold:2008zu}.

For later use, we now make the $1\leftrightarrow2$ part of the collision
term for the dilute hard population explicit.  In a pure Yang--Mills plasma,
it can be written as~\cite{Mukaida:2024jiz}
\begin{align}
 C_{1\leftrightarrow2}^{\rm LPM}[f_h](p)
 =
 \frac{(2\pi)^3}{p^2\nu_g}
 \bigg[
 &-\int_0^p \dd k\,
 \gamma_{g\leftrightarrow gg}(p;k,p-k)\,f_h(p)
 \nonumber\\
 &+\int_0^\infty \dd k\,
 2\gamma_{g\leftrightarrow gg}(p+k;p,k)\,f_h(p+k)
 \bigg],
 \label{eq:hard-lpm-collision}
\end{align}
where $\nu_g=2d_A$ is the number of gluon color and helicity degrees of
freedom, with $d_A=N_c^2-1$ for $SU(N_c)$.
The function $\gamma_{g\leftrightarrow gg}(P;k,P-k)$ describes an in-medium
splitting of a parent gluon with momentum $P$ into two daughters.
Writing the daughter momenta as $uP$ and $(1-u)P$, the leading-log
splitting function takes the form
\begin{equation}
 \gamma_{g\leftrightarrow gg}
 \bigl(P;uP,(1-u)P\bigr)
 =
 \gamma_0(P,T_s)\,\mathcal K(u),
 \label{eq:gluon-splitting-factorization}
\end{equation}
with
\begin{align}
 \gamma_0(P,T_s)
 &\equiv
 \frac{d_A C_A\alpha_s}{(2\pi)^4}
 \sqrt{\frac{P\hat q(T_s)}{2}},
 \label{eq:gluon-splitting-normalization}
 \\
 \mathcal K(u)
 &\equiv
 \frac{[1-u(1-u)]^{5/2}}
 {[u(1-u)]^{3/2}},
 \qquad 0<u<1 .
 \label{eq:bdmps-splitting-kernel}
\end{align}
Here $\gamma_0$ sets the momentum-dependent overall scale of the splitting,
while $\mathcal K(u)$ determines how the parent energy is shared between the
two daughters.

The connection with the qualitative estimate above is immediate.
Restricting the loss term in Eq.~\eqref{eq:hard-lpm-collision} to
quasi-democratic splittings, $u=\mathcal O(1)$, gives parametrically
\begin{equation}
 \Gamma_{\rm split}(P)
 \sim
 \frac{(2\pi)^3}{\nu_g P}\,
 \gamma_0(P,T_s)
 \sim
 C_A\alpha_s
 \sqrt{\frac{\hat q}{P}},
 \label{eq:gamma0-to-splitting-rate}
\end{equation}
up to logarithmic and order-one factors.
The formation-time argument captures the characteristic time scale of
the same $1\leftrightarrow2$ kernel that will be used below to describe the
full cascade.

The momentum dependence of the splitting operator determines how the cascade
develops in energy.
Since $\Gamma_{\rm split}(p)\propto p^{-1/2}$, the branching rate increases
as the particle energy decreases.  The first branching at the injection scale
$p_0$ is the bottleneck, and the cascade subsequently accelerates toward lower
momenta.

The nearly collinear kinematics instead determine the directional evolution.
At leading order, the
$C_{1\leftrightarrow2}$ collision term does not relax the angular distribution
of the cascade.  For a strictly collinear branching,
\begin{equation}
 p\hat{\bm n}
 \longrightarrow
 up\hat{\bm n}+(1-u)p\hat{\bm n},
\end{equation}
all daughter momenta remain on the same ray as the parent, while energy
conservation gives
\begin{equation}
 -p+up+(1-u)p=0.
 \label{eq:thermalization-ray-energy-invariant}
\end{equation}
The $1\leftrightarrow2$ cascade transfers energy toward lower momenta
without erasing the directional energy flow at leading order in the
collinear expansion.

The two collision terms play complementary microscopic roles.  The collinear
$C_{1\leftrightarrow2}$ cascade degrades particle energy while preserving
directional flow, whereas elastic $C_{2\leftrightarrow2}$ scattering changes
particle directions and randomizes angular correlations at the rate
$\Gamma_{\rm ang}(p)\sim\hat q/p^2$ discussed above.

We now return to the thermalization history during reheating.
The thermalization proceeds in a bottom-up manner
~\cite{Baier:2000sb,Arnold:2002zm,Kurkela:2011ti,Kurkela:2014tea,
Harigaya:2013vwa,Mukaida:2022bbo,Mukaida:2024jiz}.
The low-momentum sector can form an approximately thermal bath while the
continuously injected hard population is still cascading toward it.

The relevant cosmological comparison is between the bottleneck splitting rate
at $p_0$ and the Hubble rate. When $\Gamma_{\rm split}(p_0,T_s)\gtrsim H$, injected gluons
transfer their energy to the bath within a Hubble time. The quasi-stationary
treatment used below requires the stronger hierarchy
$\Gamma_{\rm split}(p_0,T_s)\gg H$. Even in this regime, a non-thermal hard tail
remains because inflaton decay continuously replenishes the energetic
population.

Under this quasi-stationary hierarchy, continuous injection and
splitting-induced depletion can be balanced at each momentum scale, yielding
\begin{equation}
    n_h(p)
    \equiv
    \frac{\dd n_h}{\dd\ln p}
    \sim
    \Bgfac
    \frac{m_\phi}{p}
    \frac{\Gamma_\phi}{\Gamma_{\rm split}(p)}
    n_\phi .
\label{eq:hard-tail-parametric}
\end{equation}
Here the factor $m_\phi/p$ follows from energy conservation. Each decay
injects an energy $m_\phi$, corresponding to an effective multiplicity $m_\phi/p$ at the momentum scale $p$.
Since $\Gamma_{\rm split}(p)\propto p^{-1/2}$ in the LPM regime, one obtains $n_h(p)\propto p^{-1/2}$. Using $n_h(p)=\nu_g p^3 f_h(p)/(2\pi^2)$, the hard distribution therefore
takes the characteristic form
\begin{equation}
f_h(p)\propto p^{-7/2}.
\end{equation}

This scaling describes the inertial range
\begin{equation}
    T_s \ll p \ll p_0,
    \qquad
    p_0 \simeq \frac{m_\phi}{2}.
\end{equation}
Near $p\sim p_0$, the distribution depends on the details of the
inflaton source. The stationary distribution used in our numerical analysis,
including its normalization and the treatment of the injection region, is
summarized in Appendix~\ref{app:numerics}.

The soft-bath temperature is then fixed by separating the total radiation
energy into its thermal and non-thermal components,
\begin{equation}
 \rho_R(t)=\rho_s(T_s)+\rho_h(t),
 \qquad
 \rho_s(T_s)=\frac{\pi^2}{30}g_*T_s^4.
 \label{eq:energy-closure}
\end{equation}
Here
$\rho_h(t)\equiv
\nu_g\int \frac{\dd^3p}{(2\pi)^3}\,p\,f_h(t,p)$
denotes the energy density carried by the non-thermal hard component.
This avoids assigning the energy carried by the hard cascade also to the
thermal bath.

For the bremsstrahlung calculation in Sec.~\ref{sec:bremss-finite-thermalization}, we instead condition temporarily on
the axis of a single back-to-back inflaton decay. The tagged perturbation is
then anisotropic before the final orientation average, while the ambient
plasma remains isotropic. The ensemble-level isotropy assumed above is
natural when reheating proceeds through perturbative inflaton decay, but it
may break down if preheating occurs efficiently.\footnote{For an anisotropic
distribution, the hard-loop polarization tensor cannot be reduced to a single
screening mass and may contain chromo-Weibel instabilities. The isotropic
screening prescription used below would then be insufficient
~\cite{Mrowczynski:2000ed,Mrowczynski:2004kv,Arnold:2003rq,Romatschke:2003ms,Rebhan:2004ur}.}

\section{Thermal gravitational waves under instantaneous thermalization}
\label{sec:thermal-baseline}

In Sec.~\ref{sec:cascade}, we described thermalization during reheating as
the transport of the energy injected by inflaton decay from the hard scale
$p_0\simeq m_\phi/2$ toward lower momenta and ultimately into the approximately thermal soft bath. This transport is controlled by the in-medium cascade, whose characteristic time is set by the
LPM splitting discussed there.  In the physical system this time is
finite.  To isolate the gravitational consequences of this finite thermalization time, we first establish the limiting prediction obtained when this transfer into the soft bath is taken to be instantaneous.

Within the kinetic description of Sec.~\ref{sec:cascade}, this limit amounts
to neglecting the transport time. Indeed, the quasi-stationary hard population in
Eq.~\eqref{eq:hard-tail-parametric} is proportional to
$\Gamma_{\rm split}^{-1}(p)$ and therefore disappears when the transport time
is taken to zero.  The radiation energy is then entirely contained in the equilibrated bath, and Eq.~\eqref{eq:energy-closure} reduces to the equilibrium relation between $\rho_R$ and the plasma temperature $T(t)$. The local radiation state is therefore specified by thermal equilibrium, and the injection momentum
$p_0$ no longer enters as an independent scale of that state.

To make explicit where the instantaneous-thermalization assumption enters
the gravitational-wave calculation, we first express the local production rate in terms of the energy--momentum tensor correlator without specifying
the state of the plasma.  The thermalization assumption then enters through
the state in which this correlator is evaluated.  In the instantaneous limit
described above, this state is the local equilibrium plasma characterized by
$T(t)$.

Accordingly, throughout this section we identify the soft-bath
temperature introduced in Sec.~\ref{sec:cascade} with the equilibrium
plasma temperature, $T_s(t)=T(t)$, and denote it simply by $T$.
We return to the notation $T_s$ in Sec.~\ref{sec:beyond}, where the
soft bath is distinguished from the non-thermal hard component.

\subsection{From local stress tensor to the gravitational-wave spectrum}
\label{sec:gw-general}
At leading order in the gravitational interaction, the graviton occupation number remains negligible throughout the production process. The gravitational-wave energy density then obeys
\begin{equation}
 \dot\rho_{\rm GW}+4H\rho_{\rm GW}
 =
 \gamma(t),
 \label{eq:gw-energy-balance}
\end{equation}
where $\gamma(t)$ is the GW energy emitted per unit physical volume and unit time.  The term $4H\rho_{\rm GW}$ accounts for cosmological expansion.

The production rate $\gamma(t)$ is determined by correlations of the
transverse--traceless stress tensor. Their microscopic correlation time is
much shorter than the Hubble time and the macroscopic evolution time of the
reheating background, so the production rate can be evaluated locally at each
cosmological time $t$.  We may therefore treat the plasma as stationary in a local Minkowski space.  Choosing
the graviton momentum as $\bm k=k\bm e_3$, we define the light-cone lesser Wightman correlator as 
\begin{equation}
 G_{12;12}^{<}(k,t)
 \equiv
 \int\dd^4X\,
 e^{ik(X^0-X^3)}
 \left\langle T_{12}(0)T_{12}(X)\right\rangle_t.
 \label{eq:thermal-lesser-correlator}
\end{equation}
Here $X^\mu$ denotes the relative spacetime coordinate. In this frame $T_{12}$ is a transverse--traceless stress component. The expectation value $\langle\cdots\rangle_t$ is taken in the local plasma state. Specifying this state will distinguish the instantaneous thermalization result below from the non-instantaneous thermalization result of Sec.~\ref{sec:beyond}.

The corresponding production rate per logarithmic physical graviton
momentum is~\cite{Ghiglieri:2015nfa,Ghiglieri:2020mhm}
\begin{equation}
 \mathcal{P}_{\rm GW}(k,t)
 \equiv
 \left.
 \frac{\dd\rho_{\rm GW}}
 {\dd t\,\dd\ln k}
 \right|_{\rm prod}
 =
 \frac{k^3}{\pi^2\mpl^2}
 G_{12;12}^{<}(k,t),
 \label{eq:gw-source-from-lesser}
\end{equation}
and hence
\begin{equation}
 \gamma(t)
 =
 \int_0^\infty\dd\ln k\,\mathcal{P}_{\rm GW}(k,t).
 \label{eq:gw-local-spectrum-definition}
\end{equation}
Thus, microscopic information about the plasma enters the local GW production rate through $G_{12;12}^{<}(k,t)$.

The local production rate is then accumulated over the cosmological history. A graviton observed today at frequency $f$ had
physical momentum
\begin{equation}
 k(t)=2\pi f\,\frac{a_0}{a(t)}
 \label{eq:gw-redshifting-momentum}
\end{equation}
at the emission time.
Redshifting and accumulating the local production rate
therefore gives
\begin{align}
 \Omega_{\rm GW,0}(f)
 &\equiv
 \frac{1}{\rho_{c,0}}
 \frac{\dd\rho_{\rm GW,0}}{\dd\ln f}
 \nonumber\\
 &=
 \frac{1}{\rho_{c,0}}
 \int \dd t\,
 \left(\frac{a(t)}{a_0}\right)^4
 \mathcal{P}_{\rm GW}\!\left(
  2\pi f\frac{a_0}{a(t)},t
 \right),
 \label{eq:numerical-redshift-integral}
\end{align}
where 
$\rho_{c,0}=3\mpl^2H_0^2$ is the critical energy density.  
Equation~\eqref{eq:numerical-redshift-integral} will be used throughout the following sections.  The thermalization dynamics enters through the local production rate $\mathcal P_{\rm GW}(k,t)$, while the subsequent redshift of the produced gravitons is common to the different plasma states considered
below.

\subsection{Graviton production from the thermal bath}
In the instantaneous-thermalization limit, the plasma is in local thermal equilibrium at each cosmological time.  The local stress tensor no longer retains information about how the injected energy was transported. Accordingly, the correlator introduced in Eq.~\eqref{eq:thermal-lesser-correlator} reduces to the thermal correlator,
\begin{equation}
 G_{12;12}^{<}(k,t)
 \;\longrightarrow\;
 G_{12;12}^{<,{\rm eq}}\!\left(k,T(t)\right).
 \label{eq:instantaneous-equilibrium-correlator}
\end{equation}
Thus the time dependence of the correlator is inherited from the thermal equilibrium state specified by $T(t)$.

Because the local state is in thermal equilibrium, its lesser and retarded correlators are related by the KMS relation,
\begin{equation}
 G_{12;12}^{<,{\rm eq}}(k,T)
 =
 2n_B(\hat k)\,
 \operatorname{Im}G_{12;12}^{R,{\rm eq}}(k,T),
 \qquad
 \hat k\equiv\frac{k}{T}.
 \label{eq:thermal-stress-kms}
\end{equation}
Here the retarded correlator is evaluated on the graviton light cone, $\omega=k$. The thermal GW production rate can therefore be characterized by the momentum dependence of this retarded correlator. 

To expose the momentum dependence of this thermal correlator, we factor out its overall thermal dimension and define
\begin{equation}
 \hat\eta(T,\hat k)
 \equiv
 \frac{G_{12;12}^{<,{\rm eq}}(k,T)}{2T^4}
 =
 \frac{n_B(\hat k)}{T^4}
 \operatorname{Im}G_{12;12}^{R,{\rm eq}}(k,T).
 \label{eq:eta-hat-correlator-definition}
\end{equation}
The local thermal production rate then becomes
\begin{equation}
 \mathcal{P}_{\rm GW}^{\rm th}(k,T)
 =
 \frac{2T^7}{\pi^2\mpl^2}
 \hat k^3\hat\eta(T,\hat k).
 \label{eq:thermal-log-source}
\end{equation}
Although $\hat\eta(T,\hat k)$ is defined from a single thermal correlator, the appropriate physical description depends on the momentum range. We therefore discuss separately the hard-momentum regime governed by thermal quasiparticles and the soft-momentum regime governed by hydrodynamics.

For graviton momenta at or above the screening scale $\sim gT$, the
production process involves plasma excitations whose momenta are hard. These excitations can be
described as quasiparticles, so that the light-cone correlator admits a kinetic interpretation in terms of scattering among
on-shell plasma excitations.  The hard contribution can therefore be
organized in terms of scattering of the form $
 a+b\longrightarrow c+G $.

Within the quasiparticle region, gauge-mediated reactions contain a
logarithmically enhanced contribution arising from soft spacelike
gauge-boson exchange. The exchanged momentum $Q$ satisfies
$q^0,|q|\sim gT$ and $Q^2<0$.  Medium effects are then essential for the exchange of the gauge boson, and the
corresponding propagator must be HTL resummed. Debye screening and Landau damping regulate the infrared divergence and turn it into a finite logarithm~\cite{Braaten:1989mz,Frenkel:1989br,Blaizot:2001nr,Ghiglieri:2020mhm}.  The resulting logarithmically enhanced
Standard-Model contribution is~\cite{Ghiglieri:2015nfa,Ghiglieri:2020mhm,Bernal:2024jim,Ringwald:2020ist}\footnote{
The complete leading-order thermal production rate also contains infrared-finite, non-logarithmic hard contributions at the same order~\cite{Ghiglieri:2020mhm}.  We retain only the logarithmically enhanced contribution because the present discussion focuses on the parametric dependence of the spectrum.}
\begin{equation}
 \hat\eta_{\rm HTL}^{\rm SM}(T,\hat k)
 =
 \frac{\hat k}{16\pi(e^{\hat k}-1)}
 \sum_{a=1}^{3}
 N_a\hat m_a^2(T)
 \ln\left[
 1+\frac{4\hat k^2}{\hat m_a^2(T)}
 \right],
 \label{eq:thermal-htl-rate}
\end{equation}
where $N_a=(1,3,8)$ are the gauge-boson multiplicities of
$U(1)_Y$, $SU(2)_L$, and $SU(3)_c$, and
\begin{equation}
 \hat m_a^2(T)
 \equiv
 \frac{m_a^2(T)}{T^2}
 =
 \left(
 \frac{11}{6}g_1^2(T),\,
 \frac{11}{6}g_2^2(T),\,
 2g_s^2(T)
 \right)
 \label{eq:thermal-screening-masses}
\end{equation}
are the corresponding Debye screening masses.

We next consider the opposite momentum region
$k \ll \tau_\mathrm{tr}^{-1}\sim \alpha^2 T$, where the time scale $k^{-1}$ is much longer than the transport relaxation time of the thermal plasma $\tau_\mathrm{tr}$. On such time scales, the microscopic momentum-space anisotropy associated with a stress
fluctuation is repeatedly redistributed by collisions before the graviton can resolve the individual relaxation process. The tensor correlator can therefore be replaced by its hydrodynamic, low-frequency limit.

In this limit, the leading low-frequency tensor correlator is determined by the shear viscosity, so that
\begin{equation}
 \hat\eta_\mathrm{hydro}(T,\hat k)=
 \frac{\eta_{\rm shear}(T)}{T^3}.
 \label{eq:thermal-hydro-source}
\end{equation}
At weak coupling, the dominant contribution to the shear viscosity comes from degrees of freedom whose momentum anisotropy relaxes most slowly. 

A slowly relaxing degree of freedom retains momentum anisotropy for a longer time and therefore gives a larger contribution to the shear response. Thus, above the electroweak crossover, the leading-log estimate of the shear viscosity is dominated by right-handed charged leptons, whose momentum relaxation is controlled by $U(1)_Y$ interactions~\cite{Ghiglieri:2015nfa}.  Parametrically, shear viscosity can be written as
\begin{equation}
 \frac{\eta_{\rm shear}(T)}{T^3}
 \sim
 \frac{1}{g_1^4(T)\ln[1/g_1(T)]},
 \label{eq:shear}
 \qquad
 \tau_{\rm tr}^{-1}
 \sim
 g_1^4(T)T\ln[1/g_1(T)].
\end{equation}

For $k\tau_{\rm tr}\ll1$, the graviton probes only the zero-frequency limit of the stress correlator rather than its microscopic relaxation history. This limit is characterized by the shear viscosity and is
independent of $k$ at leading order. The remaining infrared momentum dependence of the local GW production rate therefore comes entirely from the explicit graviton phase-space factor in Eq.~\eqref{eq:thermal-log-source}, yielding
\begin{equation}
\mathcal{P}_{\rm GW}^{\rm hydro}\propto k^3.
\end{equation}

The quasiparticle and hydrodynamic descriptions established above apply to separated momentum regimes of the same thermal correlator.  At weak
coupling, $\tau_{\rm tr}^{-1}\ll m_a$, leaving the intermediate momentum region
\begin{equation}
 \tau_{\rm tr}^{-1}
 \lesssim k\lesssim m_a
 \label{eq:thermal-intermediate-region}
\end{equation}
which is outside both controlled descriptions.  A systematic treatment of this region is not presently available, and perturbative expressions extended into it are known to develop pathological behavior~\cite{Ghiglieri:2015nfa,Ghiglieri:2020mhm}.  We therefore leave
the intermediate regime unresolved in the present analysis.

\subsection{Thermal gravitational-wave spectrum under instantaneous thermalization}
In this subsection, we evaluate the thermal gravitational-wave spectrum under instantaneous thermalization.

Since the production rate is governed by a higher-dimensional operator, one might expect the gravitational-wave spectrum to be dominated by production at early times rather than near the end of reheating.
However, once the cosmological expansion and the redshifting of the produced gravitational waves are taken into account, it is not obvious
which epoch gives the dominant contribution.  The relevant quantity is therefore the spectrum accumulated per logarithmic interval of the scale factor.  From
Eq.~\eqref{eq:numerical-redshift-integral},
\begin{equation}
 \frac{\dd\Omega_{\rm GW,0}^{\rm th}}{\dd\ln a}
 =
 \frac{1}{\rho_{c,0}H}
 \left(\frac{a}{a_0}\right)^4
 \mathcal P_{\rm GW}^{\rm th}(k,T).
 \label{eq:thermal-production-time-weight-general}
\end{equation}
During the matter-dominated stage, when the inflaton dominates the energy density and behaves as pressureless matter, substituting the local
thermal production rate from Eq.~\eqref{eq:thermal-log-source} gives
\begin{equation}
 \frac{\dd\Omega_{\rm GW,0}^{\rm th}}{\dd\ln a}
 \propto
 a^4H^{-1}T^7\,
 \hat k^3\hat\eta(T,\hat k)
 \propto
 a^{23/8}\,
 \hat k^3\hat\eta(T,\hat k),
 \label{eq:thermal-reheating-time-weight}
\end{equation}
In evaluating $\Omega_{\rm GW,0}^{\rm th}(f)$, $f$ is held fixed
while $a$ is integrated over.  Eq.~\eqref{eq:gw-redshifting-momentum}
then implies $k\propto a^{-1}$.  We denote the physical momentum at
the end of reheating by
$k_{\rm R}(f)\equiv 2\pi f\,a_0/a_{\rm R}$, where $a_{\rm R}$ is the
scale factor at that time.  Since $T\propto a^{-3/8}$, the corresponding
dimensionless momentum satisfies
$\hat k=\hat k_{\rm R}(f)(a/a_{\rm R})^{-5/8}$.
Substituting this relation into
Eq.~\eqref{eq:thermal-reheating-time-weight} gives, up to factors independent
of $a$ and the slow temperature dependence of the plasma
parameters~\cite{Bernal:2024jim},
\begin{equation}
 \frac{\dd\Omega_{\rm GW,0}^{\rm th}}{\dd\ln a}
 \propto
 \frac{a}{a_{\rm R}}\,
 \hat\eta\!\left(
  T,\hat k_{\rm R}(f)
  \left(\frac{a}{a_{\rm R}}\right)^{-5/8}
 \right).
\end{equation}
In the hydrodynamic regime, $\hat\eta_\mathrm{hydro}$ is independent of momentum, so the contribution increases linearly with $a/a_\mathrm{R}$. In the quasiparticle regime, neglecting the Boltzmann factor, $\hat\eta_{\rm HTL}^{\rm SM}\propto\hat k$, and the contribution therefore scales as $(a/a_{\rm R})^{3/8}$. The scale factor integral is therefore controlled by the end of reheating, and receives no power-law enhancement from
$T_{\max}/T_{\rm R}$.\footnote{
This conclusion relies on the background dynamics for which $H\propto a^{-3/2}$ and
$T\propto a^{-3/8}$.  More general reheating histories can instead produce logarithmic or power-law enhancements in $T_{\max}/T_{\rm R}$; see Ref.~\cite{Bernal:2024jim}.
}  During the subsequent radiation-dominated era, the Hubble parameter and temperature scale as $H\propto a^{-2}$, $T\propto a^{-1}$, and $\hat k=k/T$ is constant.  The contribution from the radiation-dominated era then decreases as
$a^{-1}$ and is largest at the beginning of radiation domination.

The integrand therefore increases toward $a_{\rm R}$ during
reheating and decreases after reheating. Consequently, the cosmological
integral is parametrically dominated by the end of reheating. Thus, the gravitational-wave spectrum can be estimated from the
contribution around $a\simeq a_{\rm R}$ as
\begin{align}
 \Omega_{\rm GW,0}^{\rm th}(f)
 &\simeq
 C_{\rm th}
 \left.
 \frac{\dd\Omega_{\rm GW,0}^{\rm th}}{\dd\ln a}
 \right|_{a=a_{\rm R}}
 \nonumber\\
 &=
 \frac{C_{\rm th}}{\rho_{c,0}H_{\rm R}}
 \left(\frac{a_{\rm R}}{a_0}\right)^4
 \mathcal P_{\rm GW}^{\rm th}(k_{\rm R},T_{\rm R})
 \nonumber\\
 &=
 \Omega_{\gamma,0}
 \frac{90\sqrt{10}}{\pi^5}
 \frac{C_{\rm th}}{\sqrt{g_{*,{\rm R}}}}
 \left(
  \frac{g_{*s,0}}{g_{*s,{\rm R}}}
 \right)^{4/3}
 \frac{T_{\rm R}}{\mpl}
 \hat k_{\rm R}^{\,3}(f)
 \hat\eta\!\left(T_{\rm R},\hat k_{\rm R}(f)\right),
 \label{eq:thermal-present-omega}
\end{align}
where
\begin{equation}
 \hat k_{\rm R}(f)
 \equiv
 \frac{k_{\rm R}}{T_{\rm R}}
 =
 \frac{2\pi f}{T_0}
 \left(
  \frac{g_{*s,{\rm R}}}{g_{*s,0}}
 \right)^{1/3}.
 \label{eq:thermal-khat-reheating}
\end{equation}
Here $\Omega_{\gamma,0}\equiv\rho_{\gamma,0}/\rho_{c,0}$ is the present photon energy fraction and $T_0$ is the present photon temperature, while $C_{\rm th}$ is an $\mathcal O(1)$ coefficient that accounts for the difference between the full redshift integral and
its estimate from the contribution at $a=a_{\rm R}$.

In the hydrodynamic regime,
$\hat\eta\simeq\eta_{\rm shear}/T^3$ is momentum independent [Eq.~\eqref{eq:shear}], so the gravitational-wave spectrum scales as
\begin{equation}
 \Omega_{\rm GW,0}^{\rm hydro}(f)
 \propto
 f^3.
 \label{eq:thermal-present-hydro-f3}
\end{equation}

The quasiparticle region gives a nearly universal
thermal peak. Neglecting the slowly varying logarithmic factor in Eq.~\eqref{eq:thermal-htl-rate}, its spectral dependence is
$\hat k^3 \hat\eta_{\rm HTL}^{\rm SM} \propto
\hat k^4/(e^{\hat k}-1)$, which reaches its maximum at
$\hat k_{\rm pk}\simeq3.92$.  The corresponding present-day frequency is
\begin{equation}
 f_{\rm th}^{\rm pk}
 \simeq
 \frac{3.92}{2\pi}T_0
 \left(
  \frac{g_{*s,0}}{g_{*s,{\rm R}}}
 \right)^{1/3}
 \simeq
 74~{\rm GHz}
 \left(
  \frac{g_{*s,{\rm R}}}{106.75}
 \right)^{-1/3}.
 \label{eq:thermal-peak-frequency}
\end{equation}
The logarithmic factor in the production rate, the running of the plasma parameters, and the
infrared-finite terms in the scattering rate shift the numerical peak into the
$74\text{--}80~{\rm GHz}$ range
\cite{Ghiglieri:2015nfa,Ghiglieri:2020mhm,Ringwald:2020ist}.

The universality of the spectrum follows because its shape is determined only by the combination $k/T$, while $k$ and $T$ redshift in the same way,
up to changes in the effective number of degrees of freedom.  The peak amplitude retains its dependence on the reheating temperature, and
Eq.~\eqref{eq:thermal-present-omega} gives
$\Omega_{\rm GW,0}^{\rm th,pk} \propto\Omega_{\gamma,0}\TR/\mpl$. Thus the peak frequency is set by the present photon temperature $T_0$, whereas its amplitude is controlled by the reheating temperature $\TR$.

\section{Thermal gravitational waves beyond instantaneous thermalization}
\label{sec:beyond}

As discussed in Sec.~\ref{sec:cascade}, medium-induced nearly collinear $1\leftrightarrow2$ splittings transfer the energy injected at the injection scale $p_0\simeq m_\phi/2$ toward lower momenta and ultimately into the approximately thermal soft bath characterized by $T_s$. The ensemble-averaged gluon distribution $f_g$ obeys the kinetic equation, Eq.~\eqref{eq:gluon-kinetic}. When the bottleneck splitting rate satisfies $\Gamma_{\rm split}(p_0,T_s)\gg H$, continuous injection and splitting-induced depletion balance at each momentum scale, leaving a non-thermal hard component even after the soft bath has formed. Using the decomposition introduced in Eq.~\eqref{eq:soft-hard-decomposition}, $f_g=f_s+f_h$, the hard distribution in the inertial range takes the characteristic form
\begin{equation}
 f_h(t,p)\propto p^{-7/2},
 \qquad
 T_s\ll p\ll p_0.
\end{equation}
During reheating, the full distribution remains non-thermal even when the cascade is much faster than the Hubble expansion.

In Sec.~\ref{sec:thermal-baseline}, we considered the instantaneous-thermalization limit in which the transport is taken to occur instantaneously, so that no persistent hard population remains.  We now evaluate gravitational-wave production using the finite-time distribution $f_g=f_s+f_h$ realized during reheating. The soft component reproduces the thermal contribution derived in Sec.~\ref{sec:thermal-baseline}, while the hard component gives an additional non-thermal contribution that preserves information about the injection scale $p_0$.

\subsection{Graviton production from the hard distribution}

For $k\gtrsim T_s$, the quasiparticle contribution to graviton
production can be described in terms of $2\to2$ scattering.  Let
$\mathcal P_{\rm GW}[f]$ denote the corresponding local production
rate evaluated with the phase-space distribution $f$.  For
$f=f_s+f_h$, we define the correction associated with the non-thermal
hard distribution by
\begin{equation}
 \Delta\mathcal P_{\rm GW}^{h}
 \equiv
 \mathcal P_{\rm GW}[f_s+f_h]
 -
 \mathcal P_{\rm GW}[f_s].
 \label{eq:hard-source-difference}
\end{equation}
Since the hard population is dilute, we expand this difference in
$f_h$.  Contributions involving two hard incoming particles are quadratic in $f_h$ and are neglected at linear order, in the absence of a compensating power-law enhancement at large hard momentum. For $k\gg T_s$, the purely soft contribution is Boltzmann suppressed. At linear order in $f_h$, the process contains one incoming hard gluon and one soft-bath gluon.

We evaluate this non-thermal contribution in the gluonic channel.  Since the inflaton decays into gluons, we approximate the hard population as consisting entirely of gluons and consider only $gg\to gG$.  For this
process, the cross section depends on the hard momentum only through a logarithmic factor, so the approximation of retaining only terms linear in $f_h$ is valid. In this region, the leading process is screened $t$- and $u$-channel gluon exchange:
\begin{equation}
 g_h(P)+g_s(Q)\longrightarrow G(K)+g_s(Q'),
 \qquad
 p\simeq k\gg T_s,
 \qquad
 q,q'=\mathcal O(T_s).
 \label{eq:hard-soft-process}
\end{equation}
The final-state Bose factor for the outgoing soft gluon is included in the thermal average below.

A complete Standard-Model treatment would also include hard populations of
other particle species and their corresponding scattering channels, and would require solving the coupled kinetic equations for the full
Standard-Model cascade~\cite{Mukaida:2022bbo}. These additional channels can modify the overall normalization of the spectrum. The parametric
momentum dependence studied below is already present in the gluonic channel. We leave the full SM calculation for future work.

We now estimate the graviton-production rate entering the hard--soft collisions.  This rate can be written parametrically as $\Gamma_G^h\simeq n_s\langle\sigma_G\rangle_{\rm th}$, where $n_s$ is
the thermal gluon number density and the subscript $\mathrm{th}$ denotes the thermal average over the soft-gluon distribution.

The logarithmic enhancement of $\langle\sigma_G\rangle_{\rm th}$ has the same soft-exchange origin as the
logarithm in Eq.~\eqref{eq:thermal-htl-rate}.  In particular, the $t$- and
$u$-channel gluon-exchange contributions to $gg\to gG$ are enhanced at
small momentum transfer, with
\begin{equation}
 \frac{\dd\langle\sigma_G\rangle_{\rm th}}{\dd |t|}
 \sim
 \frac{\alpha_s}{\mpl^2|t|},
\end{equation}
and the same scaling in the $u$-channel region.  At logarithmic accuracy,
the momentum-transfer integral extends from the screening scale $|t|\sim m_{s}^2$ to the Mandelstam invariant of this collision $s \sim pT_s$.\footnote{Strictly speaking, the screening mass should be evaluated using the full gluon distribution as $m_s^2\sim\alpha_s\int_{\bm p}f_g(p)/p$. Although $f_g$ contains a non-equilibrium hard distribution $f_h$, its contribution is parametrically suppressed by its small occupation number and large characteristic momentum. We therefore approximate $m_s$ by its thermal mass.}
The thermally averaged cross section therefore scales as
\begin{equation}
 \left\langle\sigma_G\right\rangle
 \sim
 \frac{\alpha_s}{\mpl^2}
 \ln\!\left(\frac{pT_s}{m_{s}^2}\right).
\end{equation}
For a thermal gluon bath, using the gluonic matrix element of
Ref.~\cite{Ghiglieri:2020mhm}, the phase-space average, including the
gluon color and helicity multiplicity and the final-state Bose factor,
gives the screened leading-logarithmic estimate
\begin{equation}
 \Gamma_G^{h}(p,T_s)
 \simeq
 \frac{N_c}{6}\frac{\alpha_sT_s^3}{\mpl^2}
 \ln\!\left(\frac{pT_s}{m_{s}^2}\right).
 \label{eq:hard-graviton-rate}
\end{equation}
The hard momentum enters this rate only through the logarithm.

The infrared-sensitive contribution involves momentum transfers down to the screening mass $m_{s}$.  For such soft exchange, with $p\gg m_{s}\sim g_sT_s$, the momentum transfer changes the hard momentum only slightly.  We therefore approximate the momentum of the produced graviton as $k\simeq p$. The energy density carried by the hard component per logarithmic momentum interval is $\dd\rho_h/\dd\ln p$. The
local production rate is then estimated as\footnote{In addition to the hard--soft $2\to2$ process, a hard gluon can radiate a graviton while undergoing one or more soft scatterings, corresponding to the collinear $1+n\leftrightarrow2+n$ scattering. We expect this contribution to be parametrically suppressed. Unlike collinear gluon emission, the transverse--traceless projection removes the collinear enhancement for graviton emission~\cite{Ghiglieri:2020mhm}, so the emission probability accumulated over $t_{\rm form}$ scales as $q_\perp^2/\mpl^2$. Using $q_\perp^2\sim\hat q(T_s)t_{\rm form}$, the corresponding rate scales as $\hat q(T_s)/\mpl^2\sim\alpha_s^2T_s^3/\mpl^2$, compared with $\Gamma_G^h\sim\alpha_sT_s^3/\mpl^2$ in Eq.~\eqref{eq:hard-graviton-rate}. It is therefore suppressed by an additional power of $\alpha_s$, with no compensating power of $p/T_s$. Since this argument is only parametric, we leave a complete nonequilibrium LPM-resummed analysis for future work.}
\begin{equation}
 \Delta\mathcal P_{\mathrm{GW}}^{h}(k,t)
 \simeq
 \left.
 \frac{\dd\rho_h}{\dd\ln p}\,
 \Gamma_G^{h}(p,T_s)
 \right|_{p\simeq k}.
 \label{eq:local-hard-conversion-estimate}
\end{equation}

The power-law momentum dependence in
Eq.~\eqref{eq:local-hard-conversion-estimate} is determined by the energy density stored in the hard component.  Using the quasi-stationary solution obtained in
Sec.~\ref{sec:cascade}, the hard energy density per logarithmic interval scales as
\begin{equation}
 \frac{\dd\rho_h}{\dd\ln p}
 \simeq
 \Bgfac\frac{\Gphi\rho_\phi}
 {\Gamma_{\rm split}(p,T_s)}
 \propto p^{1/2}.
 \label{eq:hard-energy-per-log}
\end{equation}
The inverse splitting rate is the residence time of the injected energy near
momentum $p$.  The scaling
$\Gamma_{\rm split}(p,T_s)\propto p^{-1/2}$ therefore implies that more
energy is stored at larger momenta within the quasi-stationary hard distribution.

Combining
Eq.~\eqref{eq:local-hard-conversion-estimate} and Eq.~\eqref{eq:hard-graviton-rate}
with the splitting rate obtained in Sec.~\ref{sec:cascade}, and neglecting
logarithmic corrections, we find
\begin{equation}
 \Delta\mathcal P_{\rm GW}^{h}(k,t)
 \simeq \Bgfac \frac{1}{6\alpha_s N_\mathrm{c}}
 \frac{\Gphi\rho_\phi T_s^2}{\mpl^2}
 \sqrt{\frac{k}{T_s}},
 \qquad
 T_s\ll k\ll p_0.
 \label{eq:hard-local-production-rate}
\end{equation}
The overall scale is set by the energy injection rate
$\Gphi\rho_\phi$, multiplied by the gravitational suppression
$T_s^2/\mpl^2$.  The remaining power-law momentum dependence is
$\sqrt{k/T_s}$, which originates from the momentum-dependent residence
time in the hard cascade.  Hence, up to logarithmic corrections, the local production rate scales as $\Delta\mathcal P_{\rm GW}^{h}\propto k^{1/2}$.

\subsection{Thermal gravitational-wave spectrum beyond instantaneous thermalization}
The preceding subsection determines the local hard graviton production rate $\Delta\mathcal P_{\rm GW}^{h}$. We now integrate this rate over the cosmological history. Using $\dd t=\dd\ln a/H$, Eq.~\eqref{eq:numerical-redshift-integral} gives the contribution per logarithmic interval of the scale factor as
\begin{equation}
 \frac{\dd\Omega_{\rm GW,0}^{h}}{\dd\ln a}
 =
 \frac{1}{\rho_{c,0}H}
 \left(\frac{a}{a_0}\right)^4
 \Delta\mathcal P_{\rm GW}^{h}(k,t).
 \label{eq:hard-production-time-weight}
\end{equation}
Holding $f$ fixed in the $\ln a$ integral and using the matter-dominated reheating scalings, Eq.~\eqref{eq:hard-local-production-rate} gives
\begin{equation}
 \frac{\dd\Omega_{\rm GW,0}^{h}}{\dd\ln a}
 \propto
 f^{1/2}a^{23/16},
 \label{eq:hard-reheating-time-weight}
\end{equation}
up to a logarithmic factor and the slow running of the plasma
parameters. Thus, the integral is dominated by the end of reheating, $a\simeq a_{\rm R}$, as discussed in Sec.~\ref{sec:thermal-baseline}.

We therefore estimate the full cosmological integral by evaluating the contribution at the end of reheating,
\begin{equation}
 \begin{aligned}
 \Omega_{\rm GW,0}^{h}(f)
 &\simeq
 C_h
 \left.
 \frac{\dd\Omega_{\rm GW,0}^{h}}{\dd\ln a}
 \right|_{a= a_{\rm R}}
 \\
 &=
 \frac{C_h}{\rho_{c,0}H_{\rm R}}
 \left(\frac{a_{\rm R}}{a_0}\right)^4
 \Delta\mathcal P_{\rm GW}^{h}(k_{\rm R},t_{\rm R}),
 \end{aligned}
 \label{eq:hard-endpoint-estimate}
\end{equation}
where $C_h=\mathcal O(1)$ parametrizes the
correction associated with this approximation.

Evaluating the local hard graviton production rate at the end of reheating, with
$k_{\rm R}=\TR\hat k_{\rm R}(f)$ and using
Eq.~\eqref{eq:hard-local-production-rate}, we obtain
\begin{equation}
 \Omega_{\rm GW,0}^{h}(f)
 \simeq \Omega_{\gamma,0}\frac{5\pi^{1/2}}{24\sqrt{2}}
 \frac{C_h  g_{*,{\rm R}}}
 {\alpha_s N_\mathrm{c}}
 \left(
  \frac{g_{*s,0}}{g_{*s,{\rm R}}}
 \right)^{7/6}
 \left(\frac{\TR}{\mpl}\right)^2
 \sqrt{\frac{f}{T_0}},
 \label{eq:hard-present-spectrum}
\end{equation}
up to logarithmic corrections. The $f^{1/2}$ slope therefore directly reflects the momentum dependence
of the energy stored in the finite-time hard cascade.

The hard cascade extends up to the injection momentum
$p_0=\mphi/2$, so the $f^{1/2}$ scaling extends up to its redshifted scale. Since the production is dominated by the end of reheating, the corresponding present-day frequency is
\begin{equation}
 \begin{aligned}
 f_{\rm inj}
 &\equiv
 \frac{p_0}{2\pi}\frac{T_0}{\TR}
 \left(
  \frac{g_{*s,0}}{g_{*s,{\rm R}}}
 \right)^{1/3}
 \\
 &\simeq
 9.43\times10^{9}\,{\rm Hz}\,
 \frac{\mphi}{\TR}
 \left(
  \frac{g_{*s,{\rm R}}}{106.75}
 \right)^{-1/3}.
 \end{aligned}
 \label{eq:nominal-injection-frequency}
\end{equation}

Using the quasi-stationary solution of the kinetic equation, we have evaluated graviton production from a plasma undergoing finite-time thermalization. In the instantaneous-thermalization approximation, the thermal distribution is exponentially suppressed above the thermal scale $T_s$, rendering thermal graviton production negligible in this region. Finite-time thermalization, by contrast, sustains a population of particles throughout the hard cascade, whose scattering with the thermal bath generates a non-thermal GW component scaling as $f^{1/2}$. Combining this component with the thermal-bath contribution of Sec.~\ref{sec:thermal-baseline} yields the GW spectrum including finite-time thermalization effects.

\section{Graviton bremsstrahlung beyond instantaneous thermalization}
\label{sec:bremss-finite-thermalization}

In Sec.~\ref{sec:cascade}, the inflaton source was written as an ensemble
average over randomly oriented back-to-back injections
$\mathcal I_{\hat{\bm n}}$.  The hard distribution used in Sec.~\ref{sec:beyond} is obtained after evolving this ensemble source.
Gravitational bremsstrahlung instead requires us to retain the directional
and spacetime structure of an individual injection before performing the
ensemble average.  We therefore consider a localized decay source inserted
at $x'=(t',\bm x')$ with axis $\hat{\bm n}$.  Here the localization refers
to the local decay vertex in the kinetic description; the homogeneous
reheating source is recovered after integrating over the insertion point and
averaging over the decay orientation.

Once the soft bath has formed, the dilute hard component can be treated as a
linear perturbation on top of the approximately isotropic thermal background.
We denote the corresponding linearized collision operator by
\begin{equation}
 \mathcal L_h(t)
 \equiv
 C_{1\leftrightarrow2}^{\rm LPM}(t)
 +
 C_{2\leftrightarrow2}^{\rm el}(t),
\end{equation}
and define the retarded response to a localized decay injection at
$X'=(t',\bm x')$ by
\begin{equation}
 \left[
  \partial_t
  +\hat{\bm p}\cdot\bm\nabla_{\bm x}
  -Hp\partial_p
  -\mathcal L_h(t)
 \right]
 h_{\hat{\bm n}}(X,\bm p;X')
 =
 \delta(t-t')\,
 \delta^{(3)}(\bm x-\bm x')\,
 \mathcal I_{\hat{\bm n}}(\bm p),
 \qquad
 h_{\hat{\bm n}}=0
 \quad (t<t').
 \label{eq:bremss-single-decay-response}
\end{equation}
The response $h_{\hat{\bm n}}$ contains the two primary gluons and all of
their descendants generated by subsequent branchings; it is therefore a
single-decay kinetic response rather than a tagged-particle distribution.

The ensemble-averaged hard distribution is recovered by
convolving the single-decay response with the inflaton source.  Using the
source decomposition introduced in Sec.~\ref{sec:cascade},
\begin{equation}
 f_h(t,\bm p)
 =
 \int^t \dd t'\int\dd^3x'\,
 \Bgfac\Gamma_\phi\frac{\rho_\phi(t')}{m_\phi}
 \int\frac{\dd\Omega_{\hat{\bm n}}}{4\pi}\,
 h_{\hat{\bm n}}(X,\bm p;X').
 \label{eq:hard-distribution-from-single-decay-response}
\end{equation}
For the homogeneous reheating background, the right-hand side is independent
of $\bm x$.  Equation~\eqref{eq:hard-distribution-from-single-decay-response}
makes explicit that $f_h$ is obtained after summing over decay events and
averaging over their orientations.

Sec.~\ref{sec:beyond} used this ensemble-averaged distribution to compute
graviton production in hard--soft collisions.  For gravitational
bremsstrahlung, we instead first construct the radiation amplitude generated
by the response to a single localized decay.  The two back-to-back branches
of that response are combined coherently, and the decay orientation is
averaged only after forming the radiation probability.  Spatial homogeneity
allows us to set the insertion point $\bm x'=0$ when computing this
single-decay response.

\subsection{Free-streaming limit and the origin of the soft tail}
\label{sec:free-streaming-limit}

The medium modification of the bremsstrahlung spectrum is controlled by the
microscopic evolution following each decay.  In the quasi-stationary regime
of Sec.~\ref{sec:cascade}, $\Gamma_{\rm split}(p_0)\gg H$, so this evolution
takes place on a time scale much shorter than the Hubble time.  We may
therefore treat the cosmological background as fixed when studying the local
response to a single decay.  The kinetic contribution of
$h_{\hat{\bm n}}$ to the spatial stress tensor is
\begin{equation}
 \delta T_{\hat{\bm n},{\rm kin}}^{ij}(X;X')
 =
 \nu_g\int_{\bm p}
 \frac{p^i p^j}{p}\,
 h_{\hat{\bm n}}(X,\bm p;X').
 \label{eq:bremss-single-decay-kinetic-stress}
\end{equation}
Gravitational radiation probes the transverse--traceless part of the
spacetime history of this stress.

To identify the origin of the vacuum soft behavior, we consider the
collisionless limit of the retarded response defined above.  On time scales
short compared with the Hubble time, we may also neglect the cosmological
redshift, so that the response equation reduces to
\begin{equation}
 \left(
  \partial_t+\hat{\bm p}\cdot\bm\nabla_{\bm x}
 \right)
 h_{\hat{\bm n}}(X,\bm p;X')
 =
 \delta(t-t')\delta^{(3)}(\bm x-\bm x')\,
 \mathcal I_{\hat{\bm n}}(\bm p).
 \label{eq:bremss-free-response-equation}
\end{equation}
Since the injection profile contains the two back-to-back primary gluons,
we write
\begin{equation}
 \mathcal I_{\hat{\bm n}}(\bm p)
 =
 \mathcal I_{\hat{\bm n}}^{(+)}(\bm p)
 +
 \mathcal I_{\hat{\bm n}}^{(-)}(\bm p),
 \qquad
 \mathcal I_{\hat{\bm n}}^{(\pm)}(\bm p)
 =
 \frac{(2\pi)^3}{\nu_g}
 \delta^{(3)}(\bm p\mp p_0\hat{\bm n}).
\end{equation}
For the $+$ branch, the retarded solution of
Eq.~\eqref{eq:bremss-free-response-equation} is
\begin{equation}
 h_{\hat{\bm n}}^{(+)}(X,\bm p;X')
 =
 \Theta(t-t')\,
 \delta^{(3)}
 \!\left[
  \bm x-\bm x'-\hat{\bm n}(t-t')
 \right]
 \mathcal I_{\hat{\bm n}}^{(+)}(\bm p).
 \label{eq:bremss-free-green-solution}
\end{equation}
Equation~\eqref{eq:bremss-free-green-solution} is simply the
collisionless retarded Green-function response to the localized injection:
the momentum remains $p_0\hat{\bm n}$ and the injected energy propagates
along the corresponding null ray for all future times.

Substituting this solution into
Eq.~\eqref{eq:bremss-single-decay-kinetic-stress} gives
\begin{equation}
 \delta T_{{\rm line},+}^{ij}(X;X')
 =
 p_0\,\hat n^i\hat n^j\,
 \delta^{(3)}
 \!\left[
  \bm x-\bm x'-\hat{\bm n}(t-t')
 \right]
 \Theta(t-t').
 \label{eq:bremss-ballistic-stress}
\end{equation}
For an emitted on-shell graviton with
$
 K^\mu=k(1,\hat{\bm k}),
$
the relevant quantity is the spacetime Fourier transform of the stress,
\begin{equation}
 \delta T^{ij}(K)
 \equiv
 \int\dd t\,\dd^3x\,
 e^{\ii kt-\ii\bm k\cdot\bm x}\,
 \delta T^{ij}(t,\bm x),
\end{equation}
where we have used translational invariance to set $X'=0$.
Substituting Eq.~\eqref{eq:bremss-ballistic-stress} into this definition gives
\begin{align}
 \delta T_{{\rm line},+}^{ij}(K)
 &\propto
 p_0\hat n^i\hat n^j
 \int_0^\infty\dd s\,
 e^{\ii k(1-\hat{\bm n}\cdot\hat{\bm k})s}
 \nonumber\\
 &=
 \frac{\ii p_0\hat n^i\hat n^j}
 {k(1-\hat{\bm n}\cdot\hat{\bm k})+\ii0^+}.
 \label{eq:bremss-ballistic-pole}
\end{align}
Equation~\eqref{eq:bremss-ballistic-pole} reproduces the external-leg
structure of the leading gravitational soft factor~\cite{Weinberg:1965nx}.
Its origin is particularly transparent here: the $1/k$ pole is the Fourier
transform of directional energy flow that persists along a semi-infinite
asymptotic trajectory.

The opposite daughter gives the same structure with
$\hat{\bm n}\to-\hat{\bm n}$, and the two contributions must eventually be
combined coherently.
Together with the explicit $k^3$ factor in the one-graviton phase space,
this gives
\begin{equation}
 \frac{\dd E_{\rm GW}^{\rm vac}}{\dd\ln k}
 \propto k,
 \qquad
 k\ll p_0 .
 \label{eq:bremss-vacuum-soft-scaling}
\end{equation}
The vacuum soft scaling is a direct consequence of the persistent
directional energy flow carried by the asymptotic decay products, rather
than of the local decay vertex alone.

The reheating plasma changes this asymptotic history.  Restoring the
collision terms discussed in Sec.~\ref{sec:cascade}, nearly collinear
$1\leftrightarrow2$ branchings transport the injected energy toward lower
momenta while preserving its direction at leading order in the collinear
expansion.  The first splitting of a primary gluon does not terminate the
directional stress; the daughter cascade continues to carry it collectively.
Elastic $2\leftrightarrow2$ scatterings instead relax the angular
distribution and eventually erase this directional information near the
thermal scale.  The relevant time scale for soft gravitational radiation is
thus the lifetime of the directional energy flow, rather than the lifetime
of an individual primary gluon.  The semi-infinite history underlying the
vacuum soft pole is replaced by a finite directional-memory history.

\subsection{Directional-energy transport and memory}
\label{sec:bremss-directional-energy}

We first isolate how the medium transports and erases the directional
energy injected by a single decay.  At leading order in the soft expansion,
the spatial part of the on-shell Fourier phase is
\begin{equation}
 e^{-\ii\bm k\cdot\bm x}
 =
 1-\ii\bm k\cdot\bm x+\mathcal O(k^2).
 \label{eq:bremss-soft-spatial-expansion}
\end{equation}
The leading term depends only on the spatially integrated single-decay
response while retaining its full time dependence.  Spatial integration
removes the streaming term from the kinetic equation and isolates the
evolution generated by collinear branching and elastic scattering.  We use
this reduced problem to determine how directional energy is redistributed
in momentum and how the medium eventually erases it.  We define
\begin{equation}
 \bar h_{\hat{\bm n}}(t,\bm p;X')
 \equiv
 \int\dd^3x\,
 h_{\hat{\bm n}}(X,\bm p;X').
 \label{eq:bremss-spatially-integrated-response}
\end{equation}
For the homogeneous background, this quantity is independent of the insertion
position $\bm x'$, and we henceforth write it as
$\bar h_{\hat{\bm n}}(t,\bm p;t')$.
In the local fixed-background approximation, the response depends only on the
elapsed time after the decay,
\begin{equation}
 \bar h_{\hat{\bm n}}(t,\bm p;t')
 =
 \bar h_{\hat{\bm n}}(t-t',\bm p).
 \label{eq:bremss-time-translation-local}
\end{equation}

The total energy in this response does not distinguish different directions,
whereas gravitational radiation is sourced by anisotropic energy flow.
We decompose the directional energy carried by
$\bar h_{\hat{\bm n}}$ into spherical harmonics.  We denote by
$\Omega_{\bm p}$ the solid angle specifying the momentum direction
$\hat{\bm p}\equiv\bm p/p$ and write
\begin{equation}
 \bar h_{\hat{\bm n}}(t,p,\Omega_{\bm p})
 =
 \sum_{\ell m}
 h_{\ell m}(p,t)\,
 Y_{\ell m}(\Omega_{\bm p}).
 \label{eq:bremss-angular-decomposition}
\end{equation}

To identify the part of the angular distribution relevant for gravitational
radiation, choose the graviton propagation direction $\hat{\bm k}$ as the
polar axis and denote its transverse--traceless polarization tensor by
$\epsilon_{ij}^{(\lambda)}$.  The angular dependence entering the kinetic
stress is
\begin{equation}
 \epsilon_{ij}^{(\lambda)*}\,
 \hat p^i\hat p^j ,
\end{equation}
which is a pure quadrupole on the momentum sphere,
\begin{equation}
 \epsilon_{ij}^{(\lambda)*}\,
 \hat p^i\hat p^j
 \propto
 Y_{2m}(\Omega_{\bm p}),
 \qquad
 m=\pm2 .
 \label{eq:bremss-tt-quadrupole}
\end{equation}
After inserting the spherical-harmonic expansion of
$\bar h_{\hat{\bm n}}$, the transverse--traceless stress receives a
contribution only from its $\ell=2$ angular component.
The directional memory relevant for soft gravitational radiation is the energy-weighted quadrupole carried by the single-decay response.
It is useful to distinguish where this quadrupolar energy resides in
momentum space from its momentum-integrated value.  We define
\begin{equation}
 q_{2m}(p,t)
 \equiv
 \frac{\nu_g}{(2\pi)^3}\,
 p^3 h_{2m}(p,t),
 \qquad
 \mathcal Q_{2m}(t)
 \equiv
 \int_0^\infty \dd p\,
 q_{2m}(p,t).
 \label{eq:bremss-directional-quadrupole}
\end{equation}
Here $q_{2m}(p,t)$ is the energy-weighted quadrupole at fixed momentum,
whereas $\mathcal Q_{2m}(t)$ is its momentum integral.  The factor
$p^3$ combines the radial phase-space measure with the energy weight in the
kinetic stress.

Using the spherical-harmonic expansion above, the integrated quadrupole can also be written as
\begin{equation}
 \mathcal Q_{2m}(t)
 =
 \nu_g\int_{\bm p}
 p\,Y_{2m}^{*}(\Omega_{\bm p})\,
 \bar h_{\hat{\bm n}}(t,\bm p).
 \label{eq:bremss-quadrupole-angular-form}
\end{equation}
In the local fixed-background approximation, spatial integration of the
single-decay response equation gives
\begin{equation}
 \partial_t \bar h_{\hat{\bm n}}(t,\bm p)
 =
 C_{1\leftrightarrow2}[\bar h_{\hat{\bm n}}](\bm p)
 +
 C_{2\leftrightarrow2}[\bar h_{\hat{\bm n}}](\bm p).
 \label{eq:bremss-integrated-kinetic}
\end{equation}
The collinear conservation law of Sec.~\ref{sec:cascade} can be seen
directly from the explicit gain--loss form of the
$1\leftrightarrow2$ collision term in
Eq.~\eqref{eq:hard-lpm-collision}.  For the direction-resolved response,
the same collinear kernel acts independently at each
$\Omega_{\bm p}$, since the parent and both daughters share the same
momentum direction.  Multiplying the collision term by
$pY_{2m}^{*}(\Omega_{\bm p})$ and integrating over momentum, the loss of a
parent with energy $P$ is canceled by the gain of its two daughters,
whose energies satisfy
\begin{equation}
 P=p+(P-p).
\end{equation}
The common factor $Y_{2m}^{*}(\Omega_{\bm p})$ is unchanged across the
splitting.  Hence
\begin{equation}
 \left.
 \frac{\dd\mathcal Q_{2m}}{\dd t}
 \right|_{1\leftrightarrow2}
 =
 \nu_g\int_{\bm p}
 p\,Y_{2m}^{*}(\Omega_{\bm p})\,
 C_{1\leftrightarrow2}[\bar h_{\hat{\bm n}}](\bm p)
 =0 .
 \label{eq:bremss-quadrupole-splitting-invariant}
\end{equation}
The momentum-resolved quadrupole $q_{2m}(p,t)$ changes as the cascade moves
energy toward lower momenta, while its momentum integral
$\mathcal Q_{2m}$ remains fixed in the strict collinear limit.

Elastic $2\leftrightarrow2$ scattering does not share this ray-wise
conservation law.  Sec.~\ref{sec:cascade} showed that, in the
small-momentum-transfer limit, the transverse second moment of the elastic
collision term gives
$\langle\Delta p_\perp^2(t)\rangle=\hat q\,t$.
At fixed momentum magnitude, a transverse kick changes the momentum direction:
\begin{equation}
 \delta\theta^a \simeq \frac{q_\perp^a}{p},
 \qquad a=1,2 ,
\end{equation}
where $a$ labels an orthonormal basis on the tangent plane of the momentum
sphere.  Isotropy of the bath gives
\begin{equation}
 \frac{\dd}{\dd t}
 \left\langle
 \delta\theta^a\delta\theta^b
 \right\rangle
 =
 \frac{\hat q}{2p^2}\,\delta^{ab}.
 \label{eq:bremss-angular-kick-variance}
\end{equation}

The angular part of the same gain--loss process can then be expanded in
the small displacement $\delta\bm\theta$.  The first moment vanishes by
isotropy, while the second moment gives
\begin{align}
 \left.
 C_{2\leftrightarrow2}
 [\bar h_{\hat{\bm n}}]
 \right|_{\rm ang}
 &=
 \frac12
 \frac{\dd}{\dd t}
 \left\langle
 \delta\theta^a\delta\theta^b
 \right\rangle
 \nabla_a\nabla_b
 \bar h_{\hat{\bm n}}
 \nonumber\\
 &=
 \frac{\hat q}{4p^2}
 \nabla_{\Omega_{\bm p}}^2
 \bar h_{\hat{\bm n}} .
 \label{eq:bremss-angular-diffusion}
\end{align}
Here $\nabla_{\Omega_{\bm p}}^2$ is the Laplace--Beltrami operator on the
unit momentum sphere.  This equation isolates the angular part of the
elastic Fokker--Planck operator; radial diffusion and drag are retained in the full transport calculation.
Using
$
 \nabla_{\Omega_{\bm p}}^2Y_{\ell m}
 =
 -\ell(\ell+1)Y_{\ell m},
$
the quadrupole selected by the TT stress obeys
\begin{equation}
 \left.
 C_{2\leftrightarrow2}
 [\bar h_{\hat{\bm n}}]
 \right|_{2m}^{\rm ang}
 =
 -\gamma_2(p)\,h_{2m}(p,t),
 \qquad
 \gamma_2(p)
 =
 \frac{3\hat q}{2p^2}.
 \label{eq:bremss-quadrupole-relaxation-rate}
\end{equation}
Unlike the collinear splitting term, the angular part of elastic
scattering has no corresponding quadrupole invariant.  At fixed momentum,
the $\ell=2$ component is damped as
$h_{2m}(p,t)\propto e^{-\gamma_2(p)t}$.  Equivalently,
\begin{equation}
 \left.
 \frac{\dd\mathcal Q_{2m}}{\dd t}
 \right|_{2\leftrightarrow2}^{\rm ang}
 =
 -\int_0^\infty \dd p\,
 \gamma_2(p)\,q_{2m}(p,t),
 \label{eq:bremss-quadrupole-elastic-decay}
\end{equation}
which is nonzero in general.  Elastic scattering erases the quadrupolar
anisotropy by redistributing energy over momentum directions.  The total
energy remains conserved by the elastic collisions; it is the directional
memory, rather than the energy itself, that is lost.

The fate of the directional quadrupole at a given momentum is controlled by
the competition between branching and angular relaxation.  Using
Eq.~\eqref{eq:bremss-quadrupole-relaxation-rate} together with the
branching rate derived in Sec.~\ref{sec:cascade},
\begin{equation}
 \frac{\gamma_2(p)}
 {\Gamma_{\rm split}(p)}
 \sim
 \frac{3\sqrt{D_s}}{2C_A}
 \left(\frac{T_s}{p}\right)^{3/2}
 \sim
 \mathcal O(1)
 \left(\frac{T_s}{p}\right)^{3/2}.
 \label{eq:bremss-rate-comparison}
\end{equation}
For $p\gg T_s$, branching is faster than angular relaxation.  The
$1\leftrightarrow2$ cascade then transports the quadrupolar energy toward
lower momenta before elastic scattering can erase it.  As the cascade
approaches the thermal scale, $\gamma_2$ grows relative to
$\Gamma_{\rm split}$, and elastic scattering begins to dissipate the
directional quadrupole at $p=\mathcal O(T_s)$.

The directional memory survives while the cascade transports energy from
the injection scale $p_0$ down to momenta of order $T_s$, where elastic
angular relaxation becomes effective.  Its characteristic lifetime can be
estimated from a quasi-democratic cascade with
$p_j\simeq p_0/2^j$.  Since
$\Gamma_{\rm split}^{-1}(p)\propto p^{1/2}$, the time spent in successive
branchings decreases along the cascade,
\begin{equation}
 t_{\rm mem}
 \sim
 \sum_{j=0}^{N}
 \Gamma_{\rm split}^{-1}(p_j)
 \sim
 \Gamma_{\rm split}^{-1}(p_0)
 \sum_{j=0}^{N}2^{-j/2}
 =
 \mathcal O\!\left(\Gamma_{\rm split}^{-1}(p_0)\right),
 \qquad
 p_N=\mathcal O(T_s).
 \label{eq:bremss-memory-time}
\end{equation}
The first branching is the bottleneck of this time sum, although it does
not erase the directional quadrupole: after each nearly collinear splitting,
the daughter cascade continues to carry it toward lower momentum.  Once the
cascade reaches $p=\mathcal O(T_s)$, elastic angular relaxation removes this
memory on a time scale comparable to
$\Gamma_{\rm split}^{-1}(T_s)$, which is parametrically shorter than
$\Gamma_{\rm split}^{-1}(p_0)$ for $p_0\gg T_s$.

The resulting isotropic component is incorporated into the soft-bath
background whose thermal gravitational-wave emission is described in
Sec.~\ref{sec:thermal-baseline}, and does not contribute to the
decay-conditioned transverse--traceless stress considered here.

The parametric estimate above can be sharpened using the collinear cascade
itself.  In the strict-collinear limit, a $1\leftrightarrow2$ splitting
leaves the momentum direction unchanged, so the angular harmonic carried
by each ray is a spectator of the branching dynamics.  The radial profile
$q_{2m}(p,t)$ consequently obeys the same BDMPS gain--loss equation as the
energy distribution in the cascade, with injection at $p=p_0$.  Quantifying
the directional-memory lifetime therefore reduces, at leading order, to
following this branching cascade from $p_0$ down to
$p=\mathcal O(T_s)$, where elastic scattering removes the quadrupole.

To determine the time profile quantitatively, we first isolate the radial
transport due to collinear splitting.  We temporarily neglect
$2\leftrightarrow2$ scattering and evolve the momentum-resolved quadrupole
defined in Eq.~\eqref{eq:bremss-directional-quadrupole} with the
$1\leftrightarrow2$ collision term alone.  Projecting the spatially
integrated kinetic equation onto the $\ell=2$ harmonic gives
\begin{equation}
 \partial_t q_{2m}(p,t)
 =
 \frac{\nu_g}{(2\pi)^3}p^3
 \int \dd\Omega_{\hat{\bm p}}\,
 Y_{2m}^*(\hat{\bm p})\,
 C_{1\leftrightarrow2}^{\rm LPM}
 [\bar h_{\hat{\bm n}}](t,\bm p).
 \label{eq:bremss-splitting-only-quadrupole}
\end{equation}
Strict collinearity makes the angular harmonic carried by each ray a
spectator of the branching dynamics: the collision term changes the
momentum magnitude but not its direction.  Eq.~\eqref{eq:bremss-splitting-only-quadrupole}
has the same radial evolution
as the usual energy-weighted BDMPS cascade.

Writing Eq.~\eqref{eq:bremss-splitting-only-quadrupole} in the standard
BDMPS variables makes this identification explicit.  We define
\begin{equation}
 z\equiv\frac{p}{p_0},
 \qquad
 \tau\equiv\frac{t}{t_{\rm br}},
 \qquad
 t_{\rm br}^{-1}
 \equiv
 \bar\alpha\sqrt{\frac{\hat q(T_s)}{p_0}},
 \qquad
 \bar\alpha\equiv\frac{\alpha_s C_A}{\pi},
 \label{eq:bremss-bdmps-variables}
\end{equation}
and normalize the radial quadrupole profile as
\begin{equation}
 D(z,\tau)
 \equiv
 \frac{p_0\,q_{2m}(p_0z,t)}
 {\mathcal Q_{2m}(0)},
 \qquad
 D(z,0)=\delta(1-z).
 \label{eq:bremss-bdmps-profile}
\end{equation}
For a fixed soft bath, Eq.~\eqref{eq:bremss-splitting-only-quadrupole}
becomes the standard energy-weighted BDMPS cascade equation
~\cite{Baier:1996kr,Zakharov:1996fv,Blaizot:2015jea},
\begin{equation}
 \partial_\tau D(z,\tau)
 =
 \int_0^1\dd u\,\mathcal K(u)
 \left[
 \sqrt{\frac{u}{z}}\,
 D\!\left(\frac{z}{u},\tau\right)\Theta(u-z)
 -
 \frac{u}{\sqrt z}\,D(z,\tau)
 \right].
 \label{eq:bremss-bdmps-evolution}
\end{equation}
Here $\mathcal K(u)$ is the splitting-fraction kernel introduced in
Sec.~\ref{sec:cascade}.  This equation describes the radial transport of
the directional quadrupole from $z=1$ toward the thermal scale.  Elastic
scattering, omitted in Eq.~\eqref{eq:bremss-bdmps-evolution}, removes the
quadrupole once the cascade reaches $z=\mathcal O(T_s/p_0)$.

The time profile can be exhibited analytically in a simplified version of
Eq.~\eqref{eq:bremss-bdmps-evolution}.  We replace the full splitting kernel
by
\begin{equation}
 \mathcal K_0(u)
 =
 \frac{1}{[u(1-u)]^{3/2}},
 \label{eq:bremss-simplified-bdmps-kernel}
\end{equation}
which retains the endpoint singularities and the
$\Gamma_{\rm split}(p)\propto p^{-1/2}$ scaling of the BDMPS cascade while
dropping order-one details of the splitting-fraction dependence.  We also take the
hierarchical limit $T_s/p_0\ll1$ and replace the thermal-scale sink by an
absorbing endpoint at $z=0$.  Denoting the resulting solution by $D_0$,
the initial-value problem with $D_0(z,0)=\delta(1-z)$ has the exact
solution~\cite{Blaizot:2015jea}
\begin{equation}
 D_0(z,\tau)
 =
 \frac{\tau}
 {\sqrt{z}(1-z)^{3/2}}
 \exp\!\left[
 -\frac{\pi\tau^2}{1-z}
 \right],
 \qquad
 0<z<1.
 \label{eq:bremss-perfect-sink-radial-solution}
\end{equation}
The solution gives the momentum distribution of the directional quadrupole
that remains in the resolved cascade.

Since $D_0$ is normalized to the initial quadrupole, its integral over the
resolved cascade gives the fraction of directional quadrupole that has not
yet reached the sink,
\begin{equation}
 \int_0^1\dd z\,D_0(z,\tau)
 =
 e^{-\pi\tau^2}
 =
 1-\pi\tau^2+\mathcal O(\tau^4).
 \label{eq:bremss-gaussian-memory-baseline}
\end{equation}
The absence of a term linear in $\tau$ reflects the finite development
time of the cascade.  Immediately after injection, no directional energy
has reached the infrared sink.  As the cascade moves to lower momentum,
the increasing splitting rate accelerates the transport toward the sink.
A single-rate relaxation law $e^{-\tau}$ instead loses directional memory
linearly from the initial time.  Eq.~\eqref{eq:bremss-gaussian-memory-baseline}
is an analytic illustration of
the cascade dynamics, not a relaxation envelope assumed in the
gravitational-wave calculation.

\subsection{Finite-frequency gravitational-wave response}
\label{sec:bremss-finite-frequency-resolvent}

The previous subsection kept the leading term in the spatial soft
expansion to isolate the collision-driven evolution of the directional
memory.  We now restore the spatial dependence required for a graviton
with
$
K^\mu=k(1,\hat{\bm k})
$.
Using the same local fixed-background approximation as in
Sec.~\ref{sec:free-streaming-limit}, we set the decay insertion at
$X'=0$ and define the spatial Fourier transform
\begin{equation}
 h_{\hat{\bm n}}(t,\bm k,\bm p)
 \equiv
 \int\dd^3x\,
 e^{-\ii\bm k\cdot\bm x}\,
 h_{\hat{\bm n}}(t,\bm x,\bm p;0).
 \label{eq:bremss-spatial-fourier-response}
\end{equation}
Equation~\eqref{eq:bremss-single-decay-response} then becomes
\begin{equation}
 \left[
  \partial_t
  +\ii\hat{\bm p}\cdot\bm k
  -\mathcal L_h
 \right]
 h_{\hat{\bm n}}(t,\bm k,\bm p)
 =
 \delta(t)\,
 \mathcal I_{\hat{\bm n}}(\bm p).
 \label{eq:bremss-finite-k-response}
\end{equation}
The streaming term in
Eq.~\eqref{eq:bremss-finite-k-response} restores the spatial propagation
that disappeared after the integration over position in
Sec.~\ref{sec:bremss-directional-energy}.  Its dependence on
$\hat{\bm p}$ couples different angular components of the response, while
the same branching and elastic collision terms govern the transport and
loss of directional energy in momentum space.  The temporal factor
$e^{\ii kt}$ enters separately when this kinetic response is converted
into the gravitational-wave amplitude.

Within the soft kinetic description, the gravitational-wave source is the
transverse--traceless part of the same single-decay stress used in
Sec.~\ref{sec:free-streaming-limit}.  Using
Eq.~\eqref{eq:bremss-single-decay-kinetic-stress}, its on-shell Fourier
transform is
\begin{equation}
 \delta T_{\hat{\bm n},{\rm kin}}^{ij}(K)
 =
 \nu_g
 \int_0^\infty \dd t\,e^{\ii kt}
 \int_{\bm p}
 \frac{p^i p^j}{p}\,
 h_{\hat{\bm n}}(t,\bm k,\bm p),
 \qquad
 K^\mu=k(1,\hat{\bm k}).
 \label{eq:bremss-finite-k-stress}
\end{equation}
The two back-to-back branches are contained in
$h_{\hat{\bm n}}$ and are combined coherently in
Eq.~\eqref{eq:bremss-finite-k-stress}.  The decay orientation is averaged
after forming the radiation probability.  The gravitational-wave energy
emitted per decay and per logarithmic graviton momentum is
\begin{equation}
 \frac{\dd E_{\rm GW}}{\dd\ln k}
 =
 \frac{k^3}{4\pi^2\mpl^2}
 \sum_{\lambda=\pm2}
 \int\frac{\dd\Omega_{\hat{\bm n}}}{4\pi}
 \left|
 \epsilon_{ij}^{(\lambda)*}
 \delta T_{\hat{\bm n},{\rm kin}}^{ij}(K)
 \right|^2 .
 \label{eq:bremss-finite-k-spectrum}
\end{equation}
In the collisionless limit,
Eq.~\eqref{eq:bremss-finite-k-spectrum} reproduces the soft limit of the
exact vacuum decay spectrum in Eq.~\eqref{eq:inflaton-brems-energy-spectrum}.

The deep infrared follows directly from the finite lifetime of the
directional memory established in
Sec.~\ref{sec:bremss-directional-energy}.  As $k\to0$, the spatially
Fourier-transformed response approaches the spatially integrated response
studied there, while $e^{\ii kt}\to1$ over the finite duration of the
anisotropic stress.  Since the transverse--traceless projection selects the
$\ell=2$ component, the zero-frequency limit is
\begin{equation}
 \epsilon_{ij}^{(\lambda)*}
 \delta T_{\hat{\bm n},{\rm kin}}^{ij}(K)
 \xrightarrow{k\to0}
 \epsilon_{ij}^{(\lambda)*}
 \nu_g
 \int_0^\infty \dd t
 \int_{\bm p}
 \frac{p^i p^j}{p}\,
 \bar h_{\hat{\bm n}}(t,\bm p)
 =
 \mathrm{const.}
 \label{eq:bremss-deep-ir-finite-stress}
\end{equation}
The constant in Eq.~\eqref{eq:bremss-deep-ir-finite-stress} is finite
because of the collision dynamics derived in
Sec.~\ref{sec:bremss-directional-energy}.  Collinear branching preserves
the integrated directional quadrupole,
Eq.~\eqref{eq:bremss-quadrupole-splitting-invariant}, while transporting
$q_{2m}(p,t)$ from the injection scale toward lower momentum.  Elastic
angular diffusion instead damps the quadrupole according to
Eq.~\eqref{eq:bremss-quadrupole-elastic-decay}.  The ratio
$\gamma_2/\Gamma_{\rm split}$ grows as $(T_s/p)^{3/2}$,
Eq.~\eqref{eq:bremss-rate-comparison}, so the directional energy carried
down by the cascade is efficiently randomized once it reaches
$p=\mathcal O(T_s)$.  No non-decaying $\ell=2$ component remains at late
times, and the transverse--traceless stress has a finite time integral.
Eq.~\eqref{eq:bremss-finite-k-spectrum} then gives
\begin{equation}
 \frac{\dd E_{\rm GW}}{\dd\ln k}
 \propto k^3,
 \qquad
 k t_{\rm mem}\ll1 .
 \label{eq:bremss-medium-deep-ir}
\end{equation}

At frequencies above the inverse memory time, the graviton resolves the
directional stress before it is erased by the medium.  For
$k t_{\rm mem}\gg1$, the Fourier integral in
Eq.~\eqref{eq:bremss-finite-k-stress} is controlled by times shorter than
the lifetime of the directional memory.  Collinear branching can redistribute
the injected energy in momentum during this stage, but it preserves the
directional energy along the cascade.  The leading soft stress then recovers
the same endpoint behavior as in the free-streaming limit of
Sec.~\ref{sec:free-streaming-limit},
\begin{equation}
 \epsilon_{ij}^{(\lambda)*}
 \delta T_{\hat{\bm n},{\rm kin}}^{ij}(K)
 \propto \frac{1}{k}.
 \label{eq:bremss-resolved-soft-stress}
\end{equation}
This regime also requires $k\ll p_0$, so that the graviton remains soft,
and
$
 k\,t_{\rm form}(p_0)\ll1,
$
so that it does not resolve the microscopic formation process already
coarse-grained into the LPM splitting kernel.  In this resolved-soft window,
\begin{equation}
 t_{\rm mem}^{-1}\ll k\ll p_0,
 \qquad
 k\,t_{\rm form}(p_0)\ll1,
 \label{eq:bremss-resolved-soft-window}
\end{equation}
Eq.~\eqref{eq:bremss-finite-k-spectrum} gives
\begin{equation}
 \frac{\dd E_{\rm GW}}{\dd\ln k}
 \propto k.
 \label{eq:bremss-medium-resolved-soft}
\end{equation}

The two asymptotic regimes are separated by a smooth crossover at a
frequency of order the inverse directional-memory time.  Its precise
location depends on the full time and spatial evolution of the cascade and
cannot be fixed by the parametric estimate in
Eq.~\eqref{eq:bremss-memory-time}.  We determine it by solving
Eq.~\eqref{eq:bremss-finite-k-response} with the full BDMPS branching
kernel and the leading-log Fokker--Planck collision operator.  We define
the turnover as the midpoint in logarithmic slope between the cubic and
linear asymptotes,
\begin{equation}
 \left.
 \frac{\dd\ln\!\left(\dd E_{\rm GW}/\dd\ln k\right)}
 {\dd\ln k}
 \right|_{k=k_{\rm turn}}
 =2 .
 \label{eq:bremss-turnover-definition}
\end{equation}
The numerical solution gives
\begin{equation}
 k_{\rm turn}t_{\rm br}\simeq 2.91,
 \label{eq:bremss-turnover-result}
\end{equation}
where $t_{\rm br}$ is defined in
Eq.~\eqref{eq:bremss-bdmps-variables}.  Since $t_{\rm form}(p_0)/t_{\rm br}\sim\bar\alpha$, the turnover satisfies
$k_{\rm turn}t_{\rm form}(p_0)\sim 2.91\,\bar\alpha\ll1$ at weak coupling,
consistent with the formation-time condition of
Eq.~\eqref{eq:bremss-resolved-soft-window}.

The transport calculation above is controlled in the soft regime
$k\ll p_0$.  It determines how the medium modifies the long-time
directional stress, but it does not describe the finite-recoil decay
kinematics at $k=\mathcal O(p_0)$.  The finite-recoil boundary is supplied
by the exact vacuum decay spectrum in
Eq.~\eqref{eq:inflaton-brems-energy-spectrum}.  The two descriptions have
a parametrically controlled overlap.  From
Eq.~\eqref{eq:bremss-memory-time},
$t_{\rm mem}^{-1}\sim\Gamma_{\rm split}(p_0)$, while the LPM formation-time
estimate gives
$\Gamma_{\rm split}(p_0)\sim C_A\alpha_s\,t_{\rm form}^{-1}(p_0)$.
Hence weak coupling gives
$t_{\rm mem}^{-1}\ll t_{\rm form}^{-1}(p_0)$.
In addition,
$t_{\rm form}^{-1}(p_0)/p_0
\sim\sqrt{D_s}\alpha_s(T_s/p_0)^{3/2}\ll1$
for $p_0\gg T_s$.  The common regime is
\begin{equation}
 t_{\rm mem}^{-1}
 \ll k
 \ll t_{\rm form}^{-1}(p_0)
 \ll p_0 .
 \label{eq:bremss-soft-vacuum-overlap}
\end{equation}
In this window, the medium response has already recovered the
vacuum-like soft behavior derived in
Sec.~\ref{sec:free-streaming-limit}, while the exact vacuum decay spectrum
has reached the same soft limit.  We use this common asymptotic regime to
connect the medium-modified soft spectrum to the finite-$k/p_0$ vacuum
spectrum.

\section{Numerical Results}
\label{sec:Results}
In this section, we present the gravitational-wave spectrum from perturbative reheating without assuming instantaneous thermalization. Finite-time thermalization has two consequences for the spectrum. First, the non-thermal hard component sustained by the in-medium cascade provides an additional source of gravitons through scatterings with the soft bath. This contribution gives the ultraviolet shoulder described in Eq.~\eqref{eq:hard-present-spectrum}. Second, the medium has a finite time to retain the directional energy carried by the inflaton decay products. Once this directional memory is erased by elastic scatterings, the bremsstrahlung spectrum decreases as $f^3$ in the deep infrared, as shown in Eq.~\eqref{eq:bremss-medium-deep-ir}.

For comparison, we also show the corresponding limiting spectra in which these finite-time thermalization effects are absent. For the plasma-scattering contribution, we use the instantaneous-thermalization result derived in Sec.~\ref{sec:thermal-baseline}. For inflaton-decay bremsstrahlung, we use the vacuum spectrum summarized in App.~\ref{app:inflaton-brems}. These reference spectra allow us to identify separately the modification induced by the non-thermal hard population and that induced by the finite lifetime of directional memory.

\subsection{Standard-Model-motivated benchmark}
\label{sec:results-sm}

\begin{figure}[htbp]
  \centering

  \begin{subfigure}[t]{0.49\textwidth}
    \centering
    \IfFileExists{SM_M13_TR10.pdf}{%
      \includegraphics[width=\linewidth]
      {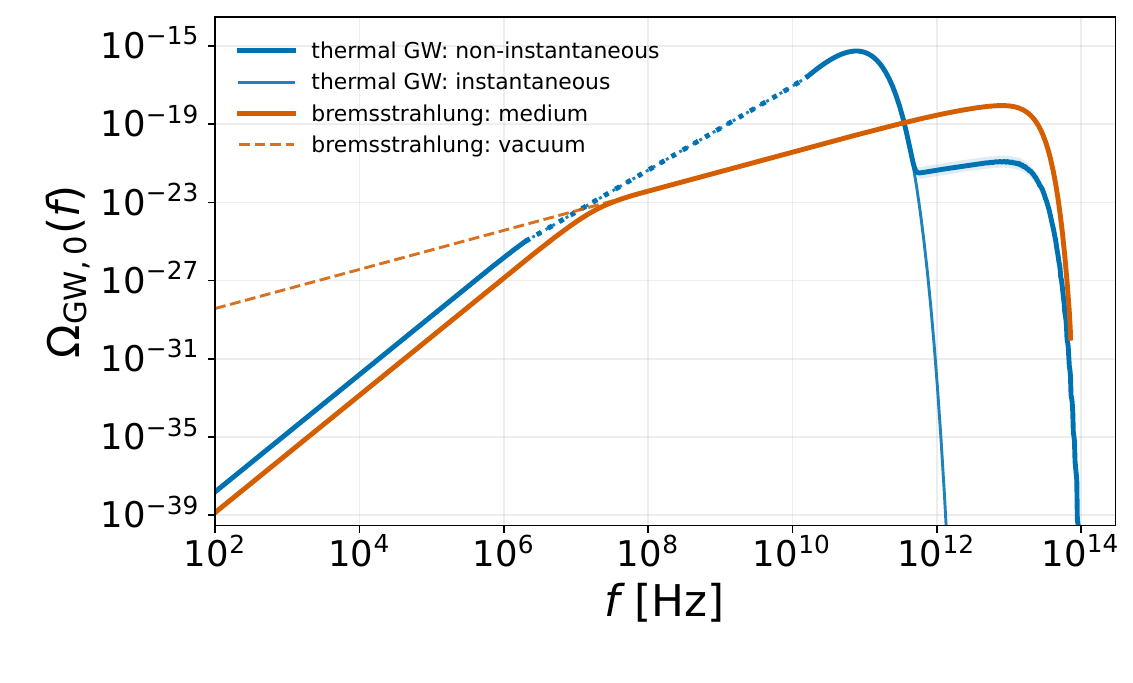}%
    }{%
      \fbox{\parbox[c][4.2cm][c]{0.92\linewidth}{\centering
      Missing figure \texttt{SM\_M13\_TR10.pdf}}}%
    }
    \caption{$m_\phi=10^{13}\,\mathrm{GeV}$, $T_R=10^{10}\,\mathrm{GeV}$.}
    \label{fig:sm-m13-tr10}
  \end{subfigure}
  \hfill
  \begin{subfigure}[t]{0.49\textwidth}
    \centering
    \IfFileExists{SM_M15_TR10.pdf}{%
      \includegraphics[width=\linewidth]
      {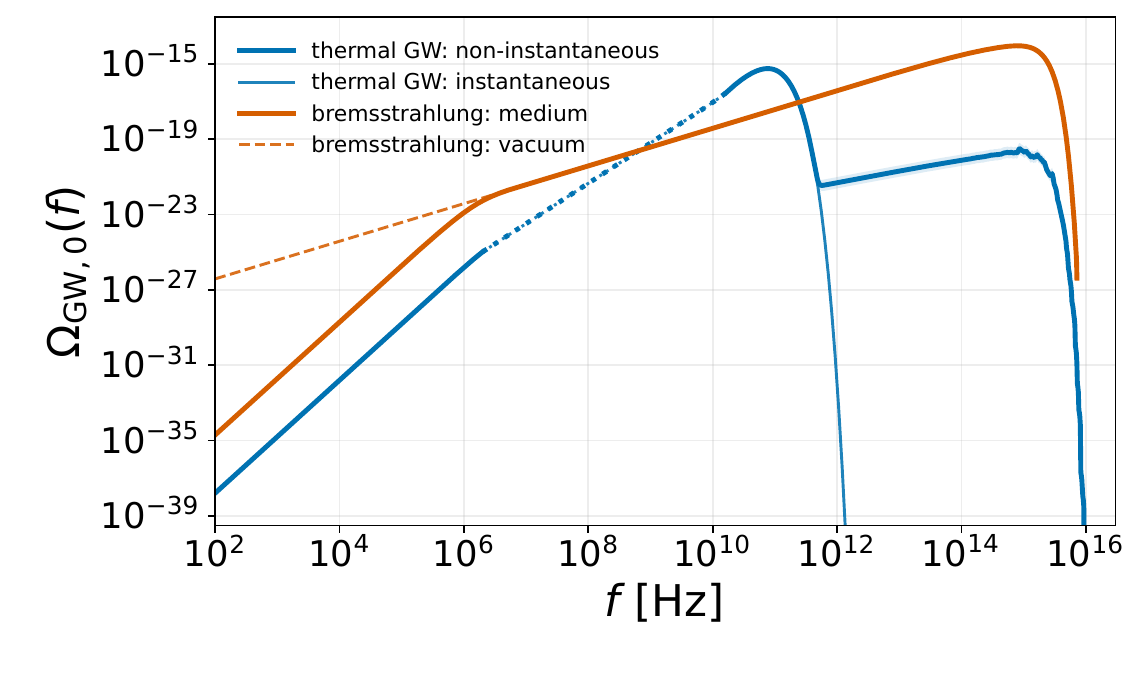}%
    }{%
      \fbox{\parbox[c][4.2cm][c]{0.92\linewidth}{\centering
      Missing figure \texttt{SM\_M15\_TR10.pdf}}}%
    }
    \caption{$m_\phi=10^{15}\,\mathrm{GeV}$, $T_R=10^{10}\,\mathrm{GeV}$.}
    \label{fig:sm-m15-tr10}
  \end{subfigure}

  \caption{
    Present-day GW spectra for the Standard-Model-motivated benchmark. The thick blue curves include the soft bath and the non-thermal hard cascade, while the thin blue curves give the instantaneous-thermalization result. The solid orange curves show the bremsstrahlung spectrum in medium matched to the vacuum result, while the dashed orange curves show only vacuum bremsstrahlung results. The dotted blue segments mark the interpolation between the hydrodynamic and hard-thermal-loop regimes. The shaded bands show the range obtained by varying the normalization of the quasi-stationary solution.}
  \label{fig:sm-comparison}
\end{figure}

We consider a Standard-Model-motivated benchmark. The thermal contribution is evaluated for the Standard Model plasma, while the non-thermal hard cascade, hard--soft scattering, and in-medium bremsstrahlung are treated in the gluonic channel, consistently with the working decay mode $\phi\to gg$. Throughout this subsection, we fix the reheating temperature to $T_R=10^{10}\,\mathrm{GeV}$ and use representative Standard Model gauge couplings appropriate to the temperatures of interest, with $\alpha_s=0.03$ for the QCD interactions governing the hard cascade. We vary the inflaton mass in order to isolate the dependence on the injection scale $p_0=m_\phi/2$.

Figure~\ref{fig:sm-comparison} shows the resulting spectra for
$m_\phi=10^{13}\,\mathrm{GeV}$ in the left panel and
$m_\phi=10^{15}\,\mathrm{GeV}$ in the right panel. In both cases, the scattering contribution develops the characteristic non-thermal ultraviolet component with
$\Omega_{\rm GW}\propto f^{1/2}$ in the hard regime (blue), while the spectrum of the graviton bremsstrahlung in medium approaches
$\Omega_{\rm GW}\propto f^3$ in the deep infrared (orange). These behaviors reproduce the analytic scalings derived in Secs.~\ref{sec:beyond} and~\ref{sec:bremss-finite-thermalization}, respectively.

Varying $m_\phi$ changes the energy of each primary decay product, $p_0=m_\phi/2$.
The relevant measure of the injected energy relative to the thermal scale is $p_0/T_R=m_\phi/(2T_R)$.
This ratio controls the separation between the thermal feature and the structures associated with the injected hard particles. For the scattering contribution, the non-thermal component extends to higher frequencies because the endpoint of the hard cascade is set by the injection momentum,
$f_{\rm inj}\propto p_0/T_R\propto m_\phi/T_R$. For bremsstrahlung, the infrared turnover instead moves toward lower frequencies. The LPM splitting rate at the injection scale decreases as $\Gamma_{\rm split}(p_0)\propto p_0^{-1/2}$, so a larger $m_\phi$ increases the directional-memory time,
$t_{\rm mem}\sim\Gamma_{\rm split}^{-1}(p_0)$, and hence decreases the corresponding turnover frequency
$f_{\rm mem}\propto t_{\rm mem}^{-1}$.

\subsection{Pure Yang--Mills results}
\label{sec:results-pureym}

\begin{figure}[htbp]
  \centering

  \begin{subfigure}[t]{0.49\textwidth}
    \centering
    \IfFileExists{PureYM_a1em2_M13_TR10.pdf}{%
      \includegraphics[width=\linewidth]
      {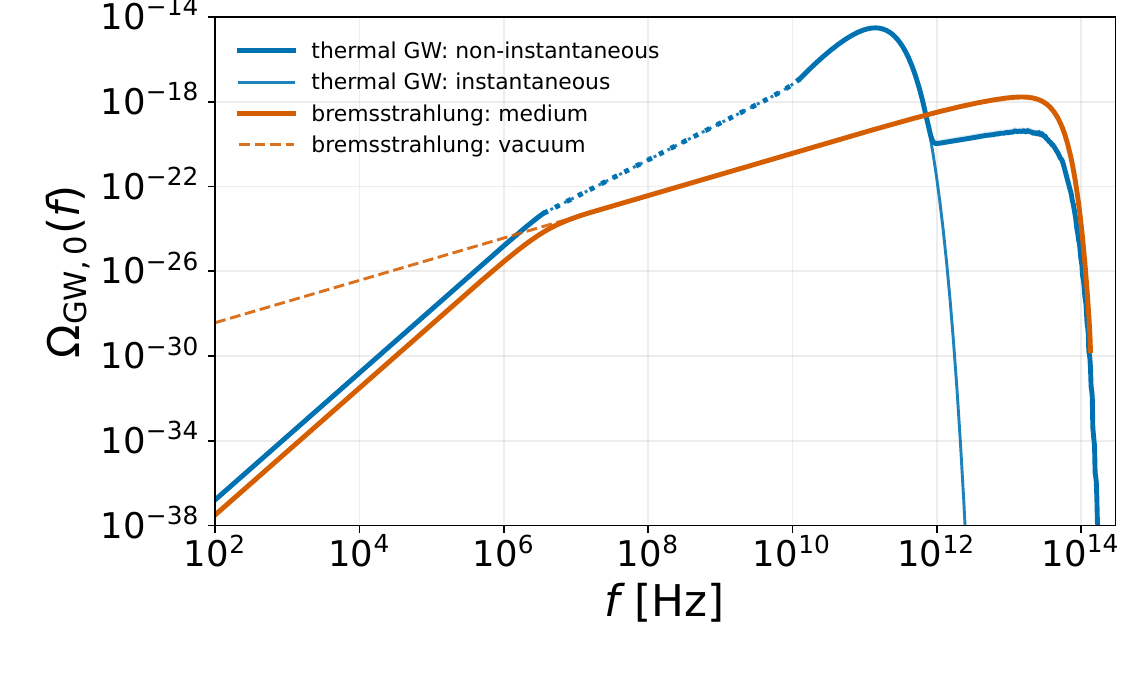}%
    }{%
      \fbox{\parbox[c][4.2cm][c]{0.92\linewidth}{\centering
      Missing figure \texttt{PureYM\_a1em2\_M13\_TR10.pdf}}}%
    }
    \caption{$\alpha_{\rm YM}=10^{-2}$.}
    \label{fig:pureym-a1em2}
  \end{subfigure}
  \hfill
  \begin{subfigure}[t]{0.49\textwidth}
    \centering
    \IfFileExists{PureYM_a1em1_M13_TR10.pdf}{%
      \includegraphics[width=\linewidth]
      {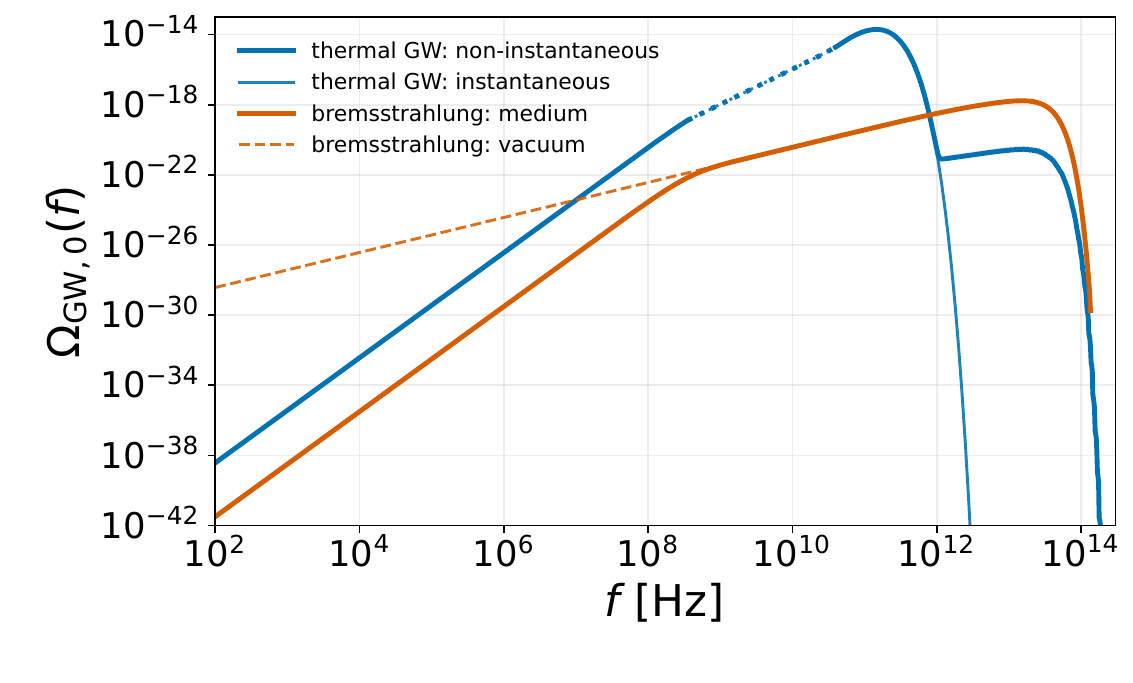}%
    }{%
      \fbox{\parbox[c][4.2cm][c]{0.92\linewidth}{\centering
      Missing figure \texttt{PureYM\_a1em1\_M13\_TR10.pdf}}}%
    }
    \caption{$\alpha_{\rm YM}=10^{-1}$.}
    \label{fig:pureym-a1em1}
  \end{subfigure}

  \caption{
  Present-day GW spectra for a dark pure $SU(3)$ Yang--Mills sector with
  $m_\phi=10^{13}\,\mathrm{GeV}$ and $T_R=10^{10}\,\mathrm{GeV}$.
  The left and right panels correspond to
  $\alpha_{\rm YM}=10^{-2}$ and $10^{-1}$, respectively.
  The axis ranges and line conventions are the same as in Fig.~\ref{fig:sm-comparison}.
  Since the weak-coupling rates are extrapolated beyond their controlled regime for $\alpha_{\rm YM}=10^{-1}$, the right panel is shown only to
  illustrate the qualitative trend toward stronger interactions.
  }
  \label{fig:pureym-comparison}
\end{figure}

We next consider reheating into a dark pure $SU(3)$ Yang--Mills plasma in order to isolate the dependence on the interaction strength. We fix
$m_\phi=10^{13}\,\mathrm{GeV}$ and
$T_R=10^{10}\,\mathrm{GeV}$, and take
$g_*=g_{*s}=2(N_c^2-1)=16$.
In this simplified setup, the plasma quantities entering the gravitational-wave production and thermalization dynamics are controlled by a single gauge coupling,
$\alpha_{\rm YM}\equiv g_{\rm YM}^2/(4\pi)$.
In particular, the screening scale, momentum broadening, LPM splitting rates, and transport quantities such as the shear viscosity are evaluated as functions of $\alpha_{\rm YM}$. We therefore vary only $\alpha_{\rm YM}$ while keeping the injection kinematics, $p_0=m_\phi/2$, fixed.

Figure~\ref{fig:pureym-comparison} shows the resulting spectra for
$\alpha_{\rm YM}=10^{-2}$ in the left panel and
$\alpha_{\rm YM}=10^{-1}$ in the right panel.
As in the Standard Model case, the spectra exhibit the same characteristic structures associated with the non-thermal hard component and with the finite directional-memory time of the bremsstrahlung source. Since $m_\phi$ and $T_R$ are fixed, the ratio
$p_0/T_R=m_\phi/(2T_R)$
is common to the two benchmarks. Accordingly, the frequency scale associated with the injected hard particles, and hence the characteristic location of the hard component, remains approximately unchanged.

The main effect of varying $\alpha_{\rm YM}$ on the scattering contribution is instead a change in its amplitude. The non-thermal hard component becomes larger as $\alpha_{\rm YM}$ decreases, consistently with the analytic behavior discussed in Sec.~\ref{sec:beyond}. A weaker interaction delays the degradation of the injected hard particles and increases their relative abundance during thermalization, thereby enhancing the gravitational-wave production sourced by the hard component.

The interaction strength also controls the directional-memory time relevant for bremsstrahlung. Increasing $\alpha_{\rm YM}$ enhances both the LPM branching and elastic angular-relaxation rates, thereby shortening the directional-memory time and shifting the corresponding infrared turnover toward higher frequencies. This trend is visible in the $\alpha_{\rm YM}=10^{-1}$ spectrum in the right panel. Since this benchmark extrapolates the weak-coupling transport treatment into a regime where it is no longer parametrically controlled, its detailed spectral shape and turnover position should be regarded as qualitative.

\section{Conclusion and Discussion}
\label{sec:conclusion}

We have studied gravitational-wave production during perturbative reheating without assuming instantaneous thermalization of the inflaton decay products.  Once an approximately thermal soft bath has formed, continuous inflaton decay still sustains a dilute non-thermal cascade
extending from the bath scale $T_s$ up to the injection scale
$p_0=\mphi/2$.  This coexistence of a thermal soft sector and a non-thermal hard cascade gives rise to two distinct gravitational-wave signatures: additional graviton production from hard--soft scatterings and a medium-induced modification of the prompt bremsstrahlung spectrum.

The first effect is additional graviton production from scatterings between
the non-thermal hard cascade and the soft bath.  The process
$g_h+g_s\to g+G$ provides a source of gravitational radiation that is absent
in the instantaneous-thermalization description.  In the inertial range,
the LPM-induced cascade with $f_h(p)\propto p^{-7/2}$ produces a broad
ultraviolet shoulder,
\begin{equation}
 \Omega_{\rm GW,0}^{h}(f)
 \propto
 f^{1/2}\times\text{slow logarithms}.
\end{equation}
The shoulder turns over near the redshifted injection scale,
$f_{\rm inj}\propto \mphi/\TR$.  The equilibrium thermal feature traces the
soft-bath scale, while the non-thermal shoulder retains information about
the hard particles injected during reheating.

The second effect is the medium modification of graviton bremsstrahlung
associated with the prompt inflaton decay.  Nearly collinear LPM branchings
transport the directional energy of the decay products toward lower
momenta without erasing it, while elastic scattering removes this
directional information near the thermal scale.  The finite lifetime of
the directional stress eliminates the vacuum soft pole and gives
\begin{equation}
 \frac{\dd E_{\rm GW}}{\dd\ln k}
 \propto
 \begin{cases}
  k^3 &\text{for}~~ k\ll k_{\rm turn},\\
  k  &\text{for}~~ k_{\rm turn}\ll k\ll p_0,
 \end{cases}
 \qquad
 k_{\rm turn}\sim\Gamma_{\rm split}(p_0),
 \label{eq:conclusion-bremss-scaling}
\end{equation}
up to an order-one factor in the turnover scale fixed by the full transport
dynamics.  Below the turnover, the graviton probes the finite duration of
the directional stress; above it, the graviton resolves the directional
energy flow before the medium erases it and the vacuum-like soft behavior
is recovered.  Matching this transport regime to the exact finite-$k/p_0$
vacuum decay spectrum gives the full sequence
$\Omega_{\rm GW,0}^{\rm brem}\propto f^3\to f\to$ the finite-$x$ decay
peak.  The infrared turnover is controlled by thermalization, whereas the
high-frequency peak is set by the decay kinematics.

The two effects have been treated separately in this work.  A natural next
step is to formulate them in terms of the full stress-tensor correlation
conditioned on a single inflaton decay.  Such a formulation would describe
the coherent bremsstrahlung response and scattering-induced radiation in a
common framework, including their overlap and interference.  A second
direction is to extend the gluonic benchmark to the full Standard Model,
including the coupled evolution of the different particle species and their
decay and scattering channels.  Finally, the present analysis assumes that
the microscopic cascade is fast compared with the Hubble expansion.  When
the two time scales become comparable, the thermalization dynamics and the
cosmological evolution must be followed simultaneously.

Taken together, these results show that the gravitational-wave spectrum can
retain information about the approach to thermal equilibrium, not only about
the properties of the thermal bath after it has formed.  The ultraviolet
shoulder remembers the hard injection scale, while the infrared turnover of
prompt bremsstrahlung reflects the finite lifetime of directional energy flow
during the cascade.  Conversely, a component that remains free streaming
would retain a persistent directional stress and hence a vacuum-like linear
tail in the deep infrared.  The low-frequency behavior can thus distinguish
complete isotropization from a reheating sector containing long-lived
free-streaming degrees of freedom.  Finite-time thermalization can imprint
distinct dynamical scales on different parts of the spectrum, allowing
high-frequency gravitational waves to probe how the hot Universe was formed.

\paragraph{Note added.---}
While this manuscript was being completed,
Ref.~\cite{Blas:2026yqb} appeared and independently identified the same
infrared effect: interactions with the surrounding medium give the
directional stress carried by the decay products a finite lifetime,
changing the bremsstrahlung spectrum from
$\dd E_{\rm GW}/\dd\ln k\propto k$ to
$\dd E_{\rm GW}/\dd\ln k\propto k^3$
at sufficiently low frequency.

Ref.~\cite{Blas:2026yqb} parametrizes the loss of the daughter-particle
stress by an exponential ansatz with a characteristic relaxation time
$t_\star$, namely $\exp(-t/t_\star)$.  In the present work, we instead follow the directional memory through
the kinetic evolution of the decay products.  Collinear LPM branching
transports the directional quadrupole toward lower momenta without erasing
it, while elastic angular diffusion removes it near the thermal scale.
Since the splitting rate increases toward lower momenta, the cascade
accelerates as it develops and the loss of directional memory becomes
steeper than a single-rate exponential.  Already in the analytically
solvable BDMPS perfect-sink limit, the memory decays as
$\exp[-\pi(t/t_{\rm br})^2]$.  The full transport calculation uses the
complete BDMPS branching kernel together with momentum-dependent elastic
relaxation and shows the same accelerated depletion more quantitatively,
fixing the finite-frequency crossover between the linear and cubic regimes.

Our analysis also includes gravitational-wave production from scatterings
between the non-thermal hard cascade and the thermal bath.  This gives the
additional hard--soft contribution studied in
Sec.~\ref{sec:beyond}, which is distinct from the modification of the
bremsstrahlung memory discussed above.
\paragraph{Acknowledgements}
We thank Julia Harz for helpful discussions at the early stage of this project. This work was supported by JSPS KAKENHI Grant Nos. JP22K14044 (K.M.), JP26K07096 (K.M.) and JP26KJ1238 (T.T.).

\appendix

\section{Graviton bremsstrahlung from inflaton decay}
\label{app:inflaton-brems}

In this appendix, we review graviton bremsstrahlung from inflaton decay
in vacuum~\cite{Nakayama:2018ptw,Huang:2019lgd,Barman:2023ymn,
Bernal:2023wus,Tokareva:2023mrt,Xu:2024fjl,Xu:2025wjq,Cline:2026jra}.
The explicit tree-level calculation makes clear that the soft-graviton
amplitude contains a $1/k$ behavior, which leads to the familiar
linear energy spectrum at small graviton momentum. Keeping the full three-body kinematics also allows us to follow the
spectrum over the entire range
\begin{equation}
x\equiv \frac{2 k}{m_\phi}=\frac{k}{p_0},
\qquad 0<x<1.
\end{equation}
The resulting vacuum spectrum provides the reference for the graviton bremsstrahlung in medium discussed in Sec.~\ref{sec:bremss-finite-thermalization}.

For definiteness, consider the representative gauge interaction
\begin{equation}
 \mathcal L_{\phi gg}
 =-\frac{g_{\phi gg}}{4}\phi F^a_{\mu\nu}\widetilde F^{a\mu\nu},
 \qquad
 \Gamma_{gg}
 =\frac{\dA g_{\phi gg}^2\mphi^3}{64\pi}.
 \label{eq:inflaton-gauge-interaction}
\end{equation}
We consider the vacuum $S$-matrix process
$\phi\to ggG$, with free asymptotic final states. Here and below, $|\mathcal M|^2$ denotes the squared matrix element
summed over final-state colors and physical polarizations.

For massless gauge bosons, crossing the result of
Ref.~\cite{Klose:2022knn} and converting to our graviton
normalization gives
\begin{equation}
 \frac{|\mathcal M_{\phi\to ggG}|^2}{|\mathcal M_{\phi\to gg}|^2
 }
 =
 \frac{1}{\mpl^2}
 \frac{1+(1-x)^4}{x^2},
 \label{eq:inflaton-brems-amplitude}
\end{equation}
where
\begin{equation}
|\mathcal M_{\phi\to gg}|^2
 =\frac{\dA g_{\phi gg}^2\mphi^4}{2}.
\end{equation}
The soft behavior is therefore explicit already at the amplitude
level:
\begin{equation}
 |\mathcal M_{\phi\to ggG}|^2
 \propto \frac{1}{x^2}
 \propto \frac{1}{k^2},
 \qquad
 \mathcal M_{\phi\to ggG}\propto\frac{1}{k},
 \qquad k\ll p_0.
 \label{eq:inflaton-brems-soft-amplitude}
\end{equation}
This is the usual gravitational soft pole discussed in
Sec.~\ref{sec:bremss-finite-thermalization}.

Including the corresponding three-body phase space and the symmetry factor for the two identical gauge bosons,
\begin{equation}
 \frac{\dd\Gamma_{\phi\to ggG}}{\dd x}
 =
 \frac{\mphi x}{512\pi^3}
 |\mathcal M_{\phi\to ggG}|^2 .
\end{equation}
Combining this with Eq.~\eqref{eq:inflaton-gauge-interaction} gives
\begin{equation}
 \frac{1}{\Gamma_{gg}}
 \frac{\dd\Gamma_{\phi\to ggG}}{\dd x}
 =\frac{\mphi^2}{16\pi^2\mpl^2}
 \frac{1+(1-x)^4}{x}.
 \label{eq:inflaton-brems-number-spectrum}
\end{equation}
Thus the $1/k$ behavior of the amplitude becomes the familiar $1/x$ soft behavior of the differential spectrum.  The
inflaton coupling and the gauge multiplicity cancel in the normalized
spectrum, while the finite-$x$ dependence encodes the full decay kinematics.

The quantity relevant for the GW energy production is the emitted
energy fraction per inflaton decay and per logarithmic graviton
momentum,
\begin{align}
 \mathcal E_{\rm brem}^{\rm vac}(x)
 &\equiv
 \frac{k}{\mphi\Gamma_{gg}}
 \frac{\dd\Gamma_{\phi\to ggG}}{\dd\ln k}
 \nonumber\\
 &=
 \frac{\mphi^2}{32\pi^2\mpl^2}
 x\left[1+(1-x)^4\right]\Theta(1-x).
 \label{eq:inflaton-brems-energy-spectrum}
\end{align}
Since $x\propto k$, the $1/x$ behavior of the differential decay
spectrum in Eq.~\eqref{eq:inflaton-brems-number-spectrum} gives
\begin{equation}
 \mathcal E_{\rm brem}^{\rm vac}(x)\propto x,
 \qquad x\ll1.
\end{equation}
Thus the vacuum bremsstrahlung energy spectrum is linear in the
soft-graviton regime, as discussed in
Sec.~\ref{sec:bremss-finite-thermalization}.
Its integrated energy fraction is
\begin{equation}
 \int_0^1\dd\ln x\,
 \mathcal E_{\rm brem}^{\rm vac}(x)
 =
 \frac{3\mphi^2}{80\pi^2\mpl^2}.
 \label{eq:inflaton-brems-integrated-energy}
\end{equation}
The corresponding local vacuum production rate is 
\begin{equation}
 \mathcal{P}_{\rm GW}^{{\rm brem},{\rm vac}}(k,t)
 =
 \Gamma_{gg}\rho_\phi(t)\,
 \mathcal E_{\rm brem}^{\rm vac}
 \!\left(\frac{2k}{\mphi}\right),
 \label{eq:inflaton-brems-local-source}
\end{equation}
The present-day gravitational wave spectrum follows from Eq.~\eqref{eq:numerical-redshift-integral} as
\begin{align}
 \Omega_{\rm GW,0}^{{\rm brem},{\rm vac}}(f)
 &=
 \frac{1}{\rho_{c,0}}
 \int \dd t\,
 \left(\frac{a(t)}{a_0}\right)^4
 \Gamma_{gg}\rho_\phi(t)\,
 \mathcal E_{\rm brem}^{\rm vac}\!\left(x_f(t)\right),
 \label{eq:inflaton-brems-cosmological-spectrum}
\end{align}
where $
 x_f(t)
 \equiv
 \frac{4\pi f}{\mphi}
 \frac{a_0}{a(t)}
$ is the graviton momentum fraction at the time of emission corresponding to the present-day frequency $f$.

To understand the vacuum spectrum analytically, we approximate the
reheating epoch as matter dominated and neglect the residual inflaton
decays after $a_R$.  Following the standard treatment of graviton
bremsstrahlung during perturbative reheating
\cite{Barman:2023ymn,Bernal:2023wus}, we define
\begin{equation}
 y\equiv\frac{f}{f_{\rm inj}}=x_f(a_R),
\end{equation}
where $f_{\rm inj}$ is the present-day frequency corresponding to
$k=p_0$ at $a=a_R$.
For the decay spectrum in
Eq.~\eqref{eq:inflaton-brems-energy-spectrum}, the redshift integral
can then be evaluated analytically as
\begin{equation}
 \Omega_{\rm GW,0}^{{\rm brem},{\rm vac}}(f)
 \simeq
 \mathcal A_{\rm brem}\,y\,\mathcal J_{\phi gg}(y),
 \label{eq:inflaton-brems-analytic-spectrum}
\end{equation}
where
\begin{equation}
 \mathcal A_{\rm brem}
 \equiv
 \Bgfac\Omega_{\gamma,0}\,
 \frac{g_{*,{\rm R}}}{48\pi^2}
 \left(\frac{g_{*s,0}}{g_{*s,{\rm R}}}\right)^{4/3}
 \left(\frac{\mphi}{\mpl}\right)^2
\end{equation}
sets the overall normalization of the spectrum, while
\begin{equation}
 \mathcal J_{\phi gg}(y)
 \equiv
 \left[
 1-6y+\frac{123}{10}y^{3/2}
 -9y^2+2y^3-\frac{3}{10}y^4
 \right]\Theta(y)\Theta(1-y)
\end{equation}
encodes the frequency dependence arising from the finite-$x$ decay kinematics and the redshift integral.
Since $\mathcal J_{\phi gg}(y)=1+\mathcal O(y)$ at small $y$,
Eq.~\eqref{eq:inflaton-brems-analytic-spectrum} gives
\begin{equation}
 \Omega_{\rm GW,0}^{{\rm brem},{\rm vac}}(f)\propto f,
 \qquad y\ll1,
\end{equation}
reproducing the linear vacuum spectrum in the soft-graviton limit.

The spectrum reaches its maximum at $y_{\rm pk}=0.5242$.
The corresponding present-day peak frequency is
\begin{equation}
 f_{\rm pk}
 \simeq
 4.94\times10^9\,{\rm Hz}\,
 \frac{\mphi}{\TR}
 \left(\frac{g_{*s,{\rm R}}}{106.75}\right)^{-1/3}.
 \label{eq:inflaton-brems-peak-frequency}
\end{equation}
The peak amplitude is
\begin{equation}
 \Omega_{\rm GW,0}^{\rm pk}
 \simeq
 4.1\times10^{-19}\,\Bgfac
 \left(\frac{\mphi}{10^{13}\,{\rm GeV}}\right)^2
 \left(\frac{g_{*,{\rm R}}}{106.75}\right)
 \left(\frac{g_{*s,{\rm R}}}{106.75}\right)^{-4/3}.
 \label{eq:inflaton-brems-peak-amplitude}
\end{equation}
These results show that the peak amplitude is independent of the inflaton decay coupling, whereas the peak frequency depends on the decay rate through $f_{\rm pk}\propto\mphi/\TR$.

In this appendix, we have obtained the standard graviton-bremsstrahlung
spectrum directly from the decay amplitude, retaining the full graviton
kinematics over $0<x<1$.  The calculation assumes vacuum propagation of
the gauge daughters and neglects their subsequent interactions with the
surrounding plasma.  In Sec.~\ref{sec:bremss-finite-thermalization}, we
include these medium effects.  The result derived here therefore provides
the reference spectrum corresponding to the absence of medium effects.

\section{Numerical implementation}
\label{app:numerics}

In this appendix, we collect the ingredients needed for the numerical
results presented in Sec.~\ref{sec:Results}. Two distinct numerical calculations enter our analysis.
For the hard--soft contribution, we evaluate the on-shell $gg\to gG$ production rate by combining the corresponding matrix element with the stationary hard distribution obtained in
Sec.~\ref{sec:cascade}.

For graviton bremsstrahlung, the relevant quantity is instead the finite-frequency response of the helicity-two directional-energy perturbation introduced in
Sec.~\ref{sec:bremss-finite-frequency-resolvent}.
Its evolution is governed by medium-induced $1\leftrightarrow2$ inelastic splittings, which transport the directional energy toward
lower momenta, and by $2\leftrightarrow2$ elastic scatterings, which relax its angular anisotropy.

\paragraph{Hard--soft scattering.}
The inertial solution in Eq.~\eqref{eq:hard-tail-parametric} is completed near
the injection endpoint by the stationary profile used for a localized gluon
source~\cite{Mukaida:2022bbo,Mukaida:2024jiz},
\begin{align}
 \nu_g f_h(p,t)
 &=\frac{4\pi^2\Bgfac\Gphi n_\phi(t)}{
 p_0^{5/2}T_s^{3/2}}\,
 \mathcal F_h(z)\,
 \Theta(p-3T_s)\Theta(p_0-p),
 \label{eq:numerical-hard-distribution}
 \\
 \mathcal F_h(z)
 &=
 \begin{cases}
  c_{\rm cas}z^{-7/2},&z\leq z_m,\\[1mm]
  c_{\rm edge}(1-z+\delta_{\rm edge})^{-1/2},&z_m<z\leq1,
 \end{cases}
 \label{eq:numerical-hard-profile}
\end{align}
where $z$ is the momentum fraction introduced in
Eq.~\eqref{eq:bremss-bdmps-variables}. For the numerical evaluation, we use $c_{\rm cas}=2.6$, $c_{\rm edge}=3.27$, $\delta_{\rm edge}=10^{-3}$, and $z_m=0.76317$. To consider the uncertainty in the quasi-stationary solution, we vary the overall normalization of the distribution over the range $
1.3 \leq c_{\rm cas} \leq 5.2.$

For numerical calculation, we evaluate $\Delta\mathcal P_{\rm GW}^{h}=\mathcal P_{gg\to gG}[f_s+f_h]-\mathcal P_{gg\to gG}[f_s]$ directly from the collision integral using the matrix elements and distributions given above. For an isotropic gluon occupation $f$, the exact $2\to2$ functional used here is
\begin{align}
 \mathcal P_{gg\to gG}[f](k,t)
 ={}&\frac{1}{2!}
 \int \prod_{a=1,2,3,K}\dd\Pi_a\,
 (2\pi)^4\delta^{(4)}(P_1+P_2-P_3-K)
 \nonumber\\
 &\times K^0\delta\!\left(\ln\frac{k}{K^0}\right) |\mathcal M_{gg\to gG}|^2
 f_1f_2(1+f_3).
 \label{eq:numerical-hard-source}
\end{align}

The color- and helicity-summed vacuum matrix element is given by~\cite{Ghiglieri:2020mhm}
\begin{equation}
|\mathcal M_{gg\to gG}|^2
 =\frac{8g_s^2 C_A d_A}{\mpl^2}
 \left(\frac{st}{u}+\frac{su}{t}+\frac{tu}{s}\right).
 \label{eq:numerical-hard-vacuum-matrix-element}
\end{equation}
The soft $t$- and $u$-channel singularities are screened using the
isotropic effective-kinetic-theory prescription of
Refs.~\cite{AbraaoYork:2014hbk,Kurkela:2018oqw},
\begin{equation}
 \frac{st}{u}
 \longrightarrow
 \frac{st}{u}
 \frac{\bm q_u^2}{\bm q_u^2+\xi_g^2m_D^2},
 \qquad
 \frac{su}{t}
 \longrightarrow
 \frac{su}{t}
 \frac{\bm q_t^2}{\bm q_t^2+\xi_g^2m_D^2},
 \qquad
 \xi_g=\frac{e^{5/6}}{\sqrt{8}},
 \label{eq:numerical-debye-screening}
\end{equation}
where $\bm q_t$ and $\bm q_u$ are the spatial momentum transfers in the $t$- and $u$-channels, respectively, in the plasma rest frame. The nonsingular $s$-channel contribution is left unchanged. Here $m_D$ is the Debye mass of the isotropic distribution used in the collision kernel. In the static limit, Eq.~\eqref{eq:numerical-debye-screening} is simply the replacement $1/\bm q^2\to1/(\bm q^2+\xi_g^2m_D^2)$.  Although $m_D$ is the usual Debye mass, the exchanged gluon carries a finite energy $\omega$, so the isotropic hard-loop propagator cannot be reduced pointwise to its static form. The factor $\xi_g=e^{5/6}/\sqrt8$ accounts for this dynamical screening at the level of the integrated longitudinal soft-momentum-transfer kernel~\cite{AbraaoYork:2014hbk,Boguslavski:2024kbd}.
Eq.~\eqref{eq:numerical-debye-screening} is the resulting screened-Born prescription used in the collision integral. We evaluate the remaining phase-space integral in Eq.~\eqref{eq:numerical-hard-source} and insert it into the scale factor integral in Eq.~\eqref{eq:numerical-redshift-integral}.

\paragraph{Finite-frequency gravitational-wave response.}The spectrum of graviton bremsstrahlung in the medium, given in
Eq.~\eqref{eq:bremss-finite-k-spectrum}, is determined by the transverse--traceless projection of
$\delta T_{\hat{\bm n},{\rm kin}}^{ij}(K)$, the on-shell Fourier transform of the single-decay stress given in Eq.~\eqref{eq:bremss-finite-k-stress}.  Evaluating this stress requires the single-decay kinetic response $h_{\hat{\bm n}}(t,\bm k,\bm p)$, which obeys the kinetic equation in
Eq.~\eqref{eq:bremss-finite-k-response}. For each $k$, we therefore solve Eq.~\eqref{eq:bremss-finite-k-response} and insert the resulting response
into Eq.~\eqref{eq:bremss-finite-k-stress}.

The only ingredient left implicit in
Sec.~\ref{sec:bremss-finite-thermalization} is the explicit form of the linearized collision operator $\mathcal L_h$ in Eq.~\eqref{eq:bremss-finite-k-response}.  For the homogeneous thermal bath whose temperature $T_s$ is held fixed during the single-decay kinetic response, the collision operator is 
\begin{equation}
 \mathcal L_h[h]
 =
 C_{1\leftrightarrow2}^{\rm LPM}[h]
 +
 C_{2\leftrightarrow2}^{\rm el}[h].
 \label{eq:numerical-explicit-Lh}
\end{equation}

The collinear branching contribution is
~\cite{Mukaida:2024jiz,Blaizot:2015jea}
\begin{equation}
 C_{1\leftrightarrow2}^{\rm LPM}[h]
 (p,\Omega_{\bm p})
 =
 \frac{1}{t_{\rm br}\sqrt z}
 \int_0^1\dd u\,\mathcal K(u)
 \left[
  u^{-5/2}
  h\!\left(\frac{p}{u},\Omega_{\bm p}\right)
  -
  u\,h(p,\Omega_{\bm p})
 \right],
 \qquad
 z\equiv\frac{p}{p_0}.
 \label{eq:numerical-forward-LPM}
\end{equation}
This is the same full-kernel BDMPS gain--loss operator as in Eq.~\eqref{eq:bremss-bdmps-evolution}, written for the single-decay kinetic response $h_{\hat{\bm n}}$.  In the strict-collinear limit, it
acts independently at fixed $\Omega_{\bm p}$. 

The elastic contribution is the leading-log Fokker--Planck operator for
the approximately isotropic soft bath at fixed
$T_s$~\cite{Iancu:2015bip}
\begin{equation}
 C_{2\leftrightarrow2}^{\rm el}[h]
 (p,\Omega_{\bm p})
 =
 \frac{\hat q(T_s)}{4}
 \left\{
  \frac{1}{p^2}\partial_p
  \left[
   p^2
   \left(
    \partial_p+\frac{1}{T_s}
   \right)h
  \right]
  +
  \frac{1}{p^2}
  \nabla_{\Omega_{\bm p}}^2h
 \right\}.
 \label{eq:numerical-forward-elastic}
\end{equation}
Its angular part is precisely
Eq.~\eqref{eq:bremss-angular-diffusion}, while radial diffusion and drag are retained in the full transport calculation.  

Equations.~\eqref{eq:numerical-explicit-Lh}--%
\eqref{eq:numerical-forward-elastic} specify the collision operator in
Eq.~\eqref{eq:bremss-finite-k-response} and close the single-decay kinetic
equation solved numerically.  Solving this equation as a function of $k$
and using Eqs.~\eqref{eq:bremss-finite-k-stress} and
\eqref{eq:bremss-finite-k-spectrum} determines the momentum scale at which the gravitational-wave spectrum from graviton bremsstrahlung crosses over from $k$ to $k^3$ behavior.

\bibliographystyle{utphys}
\bibliography{references}

\providecommand{\href}[2]{#2}\begingroup\raggedright\begin{thebibliography}{10}

\bibitem{Allahverdi:2010xz}
R.~Allahverdi, R.~Brandenberger, F.-Y. Cyr-Racine, and A.~Mazumdar, ``{Reheating in Inflationary Cosmology: Theory and Applications},'' \href{https://dx.doi.org/10.1146/annurev.nucl.012809.104511}{{\em Ann. Rev. Nucl. Part. Sci.} {\bfseries 60} (2010) 27--51}, \href{https://arxiv.org/abs/1001.2600}{{\ttfamily arXiv:1001.2600 [hep-th]}}.

\bibitem{Amin:2014eta}
M.~A. Amin, M.~P. Hertzberg, D.~I. Kaiser, and J.~Karouby, ``{Nonperturbative Dynamics Of Reheating After Inflation: A Review},'' \href{https://dx.doi.org/10.1142/S0218271815300037}{{\em Int. J. Mod. Phys. D} {\bfseries 24} (2014) 1530003}, \href{https://arxiv.org/abs/1410.3808}{{\ttfamily arXiv:1410.3808 [hep-ph]}}.

\bibitem{KawasakiKohriSugiyama:1999}
M.~Kawasaki, K.~Kohri, and N.~Sugiyama, ``{Cosmological Constraints on Late-time Entropy Production},'' \href{https://dx.doi.org/10.1103/PhysRevLett.82.4168}{{\em Phys. Rev. Lett.} {\bfseries 82} (1999) 4168}, \href{https://arxiv.org/abs/astro-ph/9811437}{{\ttfamily arXiv:astro-ph/9811437 [astro-ph]}}.

\bibitem{Kawasaki:2000en}
M.~Kawasaki, K.~Kohri, and N.~Sugiyama, ``{{MeV} scale reheating temperature and thermalization of neutrino background},'' \href{https://dx.doi.org/10.1103/PhysRevD.62.023506}{{\em Phys. Rev. D} {\bfseries 62} (2000) 023506}, \href{https://arxiv.org/abs/astro-ph/0002127}{{\ttfamily arXiv:astro-ph/0002127 [astro-ph]}}.

\bibitem{Hannestad:2004Reheating}
S.~Hannestad, ``{What is the lowest possible reheating temperature?},'' \href{https://dx.doi.org/10.1103/PhysRevD.70.043506}{{\em Phys. Rev. D} {\bfseries 70} (2004) 043506}, \href{https://arxiv.org/abs/astro-ph/0403291}{{\ttfamily arXiv:astro-ph/0403291 [astro-ph]}}.

\bibitem{deSalas:2015glj}
P.~F. de~Salas, M.~Lattanzi, G.~Mangano, G.~Miele, S.~Pastor, and O.~Pisanti, ``{Bounds on very low reheating scenarios after {Planck}},'' \href{https://dx.doi.org/10.1103/PhysRevD.92.123534}{{\em Phys. Rev. D} {\bfseries 92} no.~12, (2015) 123534}, \href{https://arxiv.org/abs/1511.00672}{{\ttfamily arXiv:1511.00672 [astro-ph.CO]}}.

\bibitem{Hasegawa:2019jsa}
T.~Hasegawa, N.~Hiroshima, K.~Kohri, R.~S.~L. Hansen, T.~Tram, and S.~Hannestad, ``{{MeV}-scale reheating temperature and thermalization of oscillating neutrinos by radiative and hadronic decays of massive particles},'' \href{https://dx.doi.org/10.1088/1475-7516/2019/12/012}{{\em JCAP} {\bfseries 12} (2019) 012}, \href{https://arxiv.org/abs/1908.10189}{{\ttfamily arXiv:1908.10189 [hep-ph]}}.

\bibitem{Barbieri:2025moq}
N.~Barbieri, T.~Brinckmann, S.~Gariazzo, M.~Lattanzi, S.~Pastor, and O.~Pisanti, ``{Current Constraints on Cosmological Scenarios with Very Low Reheating Temperatures},'' \href{https://dx.doi.org/10.1103/j5rj-dz1k}{{\em Phys. Rev. Lett.} {\bfseries 135} no.~18, (2025) 181003}, \href{https://arxiv.org/abs/2501.01369}{{\ttfamily arXiv:2501.01369 [astro-ph.CO]}}.

\bibitem{Allahverdi:2020bys}
R.~Allahverdi {\em et~al.}, ``{The First Three Seconds: a Review of Possible Expansion Histories of the Early Universe},'' \href{https://dx.doi.org/10.21105/astro.2006.16182}{{\em Open J. Astrophys.} {\bfseries 4} (2021) astro.2006.16182}, \href{https://arxiv.org/abs/2006.16182}{{\ttfamily arXiv:2006.16182 [astro-ph.CO]}}.

\bibitem{Ghiglieri:2015nfa}
J.~Ghiglieri and M.~Laine, ``{Gravitational wave background from Standard Model physics: Qualitative features},'' \href{https://dx.doi.org/10.1088/1475-7516/2015/07/022}{{\em JCAP} {\bfseries 07} (2015) 022}, \href{https://arxiv.org/abs/1504.02569}{{\ttfamily arXiv:1504.02569 [hep-ph]}}.

\bibitem{Ghiglieri:2020mhm}
J.~Ghiglieri, G.~Jackson, M.~Laine, and Y.~Zhu, ``{Gravitational wave background from Standard Model physics: Complete leading order},'' \href{https://dx.doi.org/10.1007/JHEP07(2020)092}{{\em JHEP} {\bfseries 07} (2020) 092}, \href{https://arxiv.org/abs/2004.11392}{{\ttfamily arXiv:2004.11392 [hep-ph]}}.

\bibitem{Castells-Tiestos:2022qgu}
L.~Castells-Tiestos and J.~Casalderrey-Solana, ``{Thermal emission of gravitational waves from weak to strong coupling},'' \href{https://dx.doi.org/10.1007/JHEP10(2022)049}{{\em JHEP} {\bfseries 10} (2022) 049}, \href{https://arxiv.org/abs/2202.05241}{{\ttfamily arXiv:2202.05241 [hep-th]}}.

\bibitem{Ghiglieri:2022rfp}
J.~Ghiglieri, J.~Sch{\"u}tte-Engel, and E.~Speranza, ``{Freezing-in gravitational waves},'' \href{https://dx.doi.org/10.1103/PhysRevD.109.023538}{{\em Phys. Rev. D} {\bfseries 109} no.~2, (2024) 023538}, \href{https://arxiv.org/abs/2211.16513}{{\ttfamily arXiv:2211.16513 [hep-ph]}}.

\bibitem{Muia:2023wru}
F.~Muia, F.~Quevedo, A.~Schachner, and G.~Villa, ``{Testing BSM physics with gravitational waves},'' \href{https://dx.doi.org/10.1088/1475-7516/2023/09/006}{{\em JCAP} {\bfseries 09} (2023) 006}, \href{https://arxiv.org/abs/2303.01548}{{\ttfamily arXiv:2303.01548 [hep-ph]}}.

\bibitem{Drewes:2023oxg}
M.~Drewes, Y.~Georis, J.~Klaric, and P.~Klose, ``{Upper bound on thermal gravitational wave backgrounds from hidden sectors},'' \href{https://dx.doi.org/10.1088/1475-7516/2024/06/073}{{\em JCAP} {\bfseries 06} (2024) 073}, \href{https://arxiv.org/abs/2312.13855}{{\ttfamily arXiv:2312.13855 [hep-ph]}}.

\bibitem{Ghiglieri:2024ghm}
J.~Ghiglieri, M.~Laine, J.~Sch{\"u}tte-Engel, and E.~Speranza, ``{Double-graviton production from Standard Model plasma},'' \href{https://dx.doi.org/10.1088/1475-7516/2024/04/062}{{\em JCAP} {\bfseries 04} (2024) 062}, \href{https://arxiv.org/abs/2401.08766}{{\ttfamily arXiv:2401.08766 [hep-ph]}}.

\bibitem{Ringwald:2020ist}
A.~Ringwald, J.~Sch{\"u}tte-Engel, and C.~Tamarit, ``{Gravitational Waves as a Big Bang Thermometer},'' \href{https://dx.doi.org/10.1088/1475-7516/2021/03/054}{{\em JCAP} {\bfseries 03} (2021) 054}, \href{https://arxiv.org/abs/2011.04731}{{\ttfamily arXiv:2011.04731 [hep-ph]}}.

\bibitem{Klose:2022knn}
P.~Klose, M.~Laine, and S.~Procacci, ``{Gravitational wave background from non-Abelian reheating after axion-like inflation},'' \href{https://dx.doi.org/10.1088/1475-7516/2022/05/021}{{\em JCAP} {\bfseries 05} (2022) 021}, \href{https://arxiv.org/abs/2201.02317}{{\ttfamily arXiv:2201.02317 [hep-ph]}}.

\bibitem{Bernal:2024jim}
N.~Bernal and Y.~Xu, ``{Thermal gravitational waves during reheating},'' \href{https://dx.doi.org/10.1007/JHEP01(2025)137}{{\em JHEP} {\bfseries 01} (2025) 137}, \href{https://arxiv.org/abs/2410.21385}{{\ttfamily arXiv:2410.21385 [hep-ph]}}.

\bibitem{Nakayama:2018ptw}
K.~Nakayama and Y.~Tang, ``{Stochastic Gravitational Waves from Particle Origin},'' \href{https://dx.doi.org/10.1016/j.physletb.2018.11.023}{{\em Phys. Lett. B} {\bfseries 788} (2019) 341--346}, \href{https://arxiv.org/abs/1810.04975}{{\ttfamily arXiv:1810.04975 [hep-ph]}}. [Erratum: Phys.Lett.B 839, 137787 (2023)].

\bibitem{Huang:2019lgd}
D.~Huang and L.~Yin, ``{Stochastic Gravitational Waves from Inflaton Decays},'' \href{https://dx.doi.org/10.1103/PhysRevD.100.043538}{{\em Phys. Rev. D} {\bfseries 100} no.~4, (2019) 043538}, \href{https://arxiv.org/abs/1905.08510}{{\ttfamily arXiv:1905.08510 [hep-ph]}}.

\bibitem{Barman:2023ymn}
B.~Barman, N.~Bernal, Y.~Xu, and {\'O}.~Zapata, ``{Gravitational wave from graviton Bremsstrahlung during reheating},'' \href{https://dx.doi.org/10.1088/1475-7516/2023/05/019}{{\em JCAP} {\bfseries 05} (2023) 019}, \href{https://arxiv.org/abs/2301.11345}{{\ttfamily arXiv:2301.11345 [hep-ph]}}.

\bibitem{Barman:2023rpg}
B.~Barman, N.~Bernal, Y.~Xu, and {\'O}.~Zapata, ``{Bremsstrahlung-induced gravitational waves in monomial potentials during reheating},'' \href{https://dx.doi.org/10.1103/PhysRevD.108.083524}{{\em Phys. Rev. D} {\bfseries 108} no.~8, (2023) 083524}, \href{https://arxiv.org/abs/2305.16388}{{\ttfamily arXiv:2305.16388 [hep-ph]}}.

\bibitem{Jiang:2024akb}
Y.~Jiang and T.~Suyama, ``{Spectrum of high-frequency gravitational waves from graviton bremsstrahlung by the decay of inflaton: case with polynomial potential},'' \href{https://dx.doi.org/10.1088/1475-7516/2025/02/041}{{\em JCAP} {\bfseries 02} (2025) 041}, \href{https://arxiv.org/abs/2410.11175}{{\ttfamily arXiv:2410.11175 [astro-ph.CO]}}.

\bibitem{Bernal:2023wus}
N.~Bernal, S.~Cl{\'e}ry, Y.~Mambrini, and Y.~Xu, ``{Probing reheating with graviton bremsstrahlung},'' \href{https://dx.doi.org/10.1088/1475-7516/2024/01/065}{{\em JCAP} {\bfseries 01} (2024) 065}, \href{https://arxiv.org/abs/2311.12694}{{\ttfamily arXiv:2311.12694 [hep-ph]}}.

\bibitem{Tokareva:2023mrt}
A.~Tokareva, ``{Gravitational waves from inflaton decay and bremsstrahlung},'' \href{https://dx.doi.org/10.1016/j.physletb.2024.138695}{{\em Phys. Lett. B} {\bfseries 853} (2024) 138695}, \href{https://arxiv.org/abs/2312.16691}{{\ttfamily arXiv:2312.16691 [hep-ph]}}.

\bibitem{Xu:2024fjl}
Y.~Xu, ``{Ultra-high frequency gravitational waves from scattering, Bremsstrahlung and decay during reheating},'' \href{https://dx.doi.org/10.1007/JHEP10(2024)174}{{\em JHEP} {\bfseries 10} (2024) 174}, \href{https://arxiv.org/abs/2407.03256}{{\ttfamily arXiv:2407.03256 [hep-ph]}}.

\bibitem{Xu:2025wjq}
X.-J. Xu, Y.~Xu, Q.~Yin, and J.~Zhu, ``{Full-spectrum analysis of gravitational wave production from inflation to reheating},'' \href{https://dx.doi.org/10.1007/JHEP10(2025)141}{{\em JHEP} {\bfseries 10} (2025) 141}, \href{https://arxiv.org/abs/2505.08868}{{\ttfamily arXiv:2505.08868 [hep-ph]}}.

\bibitem{Cline:2026jra}
J.~M. Cline and Y.~Xu, ``{Irreducible Graviton Floor from Reheating},'' \href{https://arxiv.org/abs/2605.16201}{{\ttfamily arXiv:2605.16201 [hep-ph]}}.

\bibitem{Ema:2020ggo}
Y.~Ema, R.~Jinno, and K.~Nakayama, ``{High-frequency Graviton from Inflaton Oscillation},'' \href{https://dx.doi.org/10.1088/1475-7516/2020/09/015}{{\em JCAP} {\bfseries 09} (2020) 015}, \href{https://arxiv.org/abs/2006.09972}{{\ttfamily arXiv:2006.09972 [astro-ph.CO]}}.

\bibitem{Weinberg:1965nx}
S.~Weinberg, ``{Infrared photons and gravitons},'' \href{https://dx.doi.org/10.1103/PhysRev.140.B516}{{\em Phys. Rev.} {\bfseries 140} (1965) B516--B524}.

\bibitem{Harigaya:2013vwa}
K.~Harigaya and K.~Mukaida, ``{Thermalization after/during Reheating},'' \href{https://dx.doi.org/10.1007/JHEP05(2014)006}{{\em JHEP} {\bfseries 05} (2014) 006}, \href{https://arxiv.org/abs/1312.3097}{{\ttfamily arXiv:1312.3097 [hep-ph]}}.

\bibitem{Mukaida:2015ria}
K.~Mukaida and M.~Yamada, ``{Thermalization Process after Inflation and Effective Potential of Scalar Field},'' \href{https://dx.doi.org/10.1088/1475-7516/2016/02/003}{{\em JCAP} {\bfseries 02} (2016) 003}, \href{https://arxiv.org/abs/1506.07661}{{\ttfamily arXiv:1506.07661 [hep-ph]}}.

\bibitem{Mukaida:2022bbo}
K.~Mukaida and M.~Yamada, ``{Cascades of high-energy SM particles in the primordial thermal plasma},'' \href{https://dx.doi.org/10.1007/JHEP10(2022)116}{{\em JHEP} {\bfseries 10} (2022) 116}, \href{https://arxiv.org/abs/2208.11708}{{\ttfamily arXiv:2208.11708 [hep-ph]}}.

\bibitem{Mukaida:2024jiz}
K.~Mukaida and M.~Yamada, ``{Perturbative reheating and thermalization of pure Yang-Mills plasma},'' \href{https://dx.doi.org/10.1007/JHEP05(2024)174}{{\em JHEP} {\bfseries 05} (2024) 174}, \href{https://arxiv.org/abs/2402.14054}{{\ttfamily arXiv:2402.14054 [hep-ph]}}.

\bibitem{Mehtar-Tani:2022zwf}
Y.~Mehtar-Tani, S.~Schlichting, and I.~Soudi, ``{Jet thermalization in QCD kinetic theory},'' \href{https://dx.doi.org/10.1007/JHEP05(2023)091}{{\em JHEP} {\bfseries 05} (2023) 091}, \href{https://arxiv.org/abs/2209.10569}{{\ttfamily arXiv:2209.10569 [hep-ph]}}.

\bibitem{Schlichting:2020lef}
S.~Schlichting and I.~Soudi, ``{Medium-induced fragmentation and equilibration of highly energetic partons},'' \href{https://dx.doi.org/10.1007/JHEP07(2021)077}{{\em JHEP} {\bfseries 07} (2021) 077}, \href{https://arxiv.org/abs/2008.04928}{{\ttfamily arXiv:2008.04928 [hep-ph]}}.

\bibitem{Harigaya:2019tzu}
K.~Harigaya, K.~Mukaida, and M.~Yamada, ``{Dark Matter Production during the Thermalization Era},'' \href{https://dx.doi.org/10.1007/JHEP07(2019)059}{{\em JHEP} {\bfseries 07} (2019) 059}, \href{https://arxiv.org/abs/1901.11027}{{\ttfamily arXiv:1901.11027 [hep-ph]}}.

\bibitem{Turner:1983he}
M.~S. Turner, ``{Coherent Scalar Field Oscillations in an Expanding Universe},'' \href{https://dx.doi.org/10.1103/PhysRevD.28.1243}{{\em Phys. Rev. D} {\bfseries 28} (1983) 1243}.

\bibitem{Shtanov:1994ce}
Y.~Shtanov, J.~H. Traschen, and R.~H. Brandenberger, ``{Universe reheating after inflation},'' \href{https://dx.doi.org/10.1103/PhysRevD.51.5438}{{\em Phys. Rev. D} {\bfseries 51} (1995) 5438--5455}, \href{https://arxiv.org/abs/hep-ph/9407247}{{\ttfamily arXiv:hep-ph/9407247}}.

\bibitem{Kofman:1994rk}
L.~Kofman, A.~D. Linde, and A.~A. Starobinsky, ``{Reheating after inflation},'' \href{https://dx.doi.org/10.1103/PhysRevLett.73.3195}{{\em Phys. Rev. Lett.} {\bfseries 73} (1994) 3195--3198}, \href{https://arxiv.org/abs/hep-th/9405187}{{\ttfamily arXiv:hep-th/9405187}}.

\bibitem{Kofman:1997yn}
L.~Kofman, A.~D. Linde, and A.~A. Starobinsky, ``{Towards the theory of reheating after inflation},'' \href{https://dx.doi.org/10.1103/PhysRevD.56.3258}{{\em Phys. Rev. D} {\bfseries 56} (1997) 3258--3295}, \href{https://arxiv.org/abs/hep-ph/9704452}{{\ttfamily arXiv:hep-ph/9704452}}.

\bibitem{Mukaida:2012qn}
K.~Mukaida and K.~Nakayama, ``{Dynamics of oscillating scalar field in thermal environment},'' \href{https://dx.doi.org/10.1088/1475-7516/2013/01/017}{{\em JCAP} {\bfseries 01} (2013) 017}, \href{https://arxiv.org/abs/1208.3399}{{\ttfamily arXiv:1208.3399 [hep-ph]}}.

\bibitem{Mukaida:2012bz}
K.~Mukaida and K.~Nakayama, ``{Dissipative Effects on Reheating after Inflation},'' \href{https://dx.doi.org/10.1088/1475-7516/2013/03/002}{{\em JCAP} {\bfseries 03} (2013) 002}, \href{https://arxiv.org/abs/1212.4985}{{\ttfamily arXiv:1212.4985 [hep-ph]}}.

\bibitem{Drewes:2013iaa}
M.~Drewes and J.~U. Kang, ``{The Kinematics of Cosmic Reheating},'' \href{https://dx.doi.org/10.1016/j.nuclphysb.2013.07.009}{{\em Nucl. Phys. B} {\bfseries 875} (2013) 315--350}, \href{https://arxiv.org/abs/1305.0267}{{\ttfamily arXiv:1305.0267 [hep-ph]}}. [Erratum: Nucl.Phys.B 888, 284--286 (2014)].

\bibitem{Albrecht:1982mp}
A.~Albrecht, P.~J. Steinhardt, M.~S. Turner, and F.~Wilczek, ``{Reheating an Inflationary Universe},'' \href{https://dx.doi.org/10.1103/PhysRevLett.48.1437}{{\em Phys. Rev. Lett.} {\bfseries 48} (1982) 1437}.

\bibitem{Chung:1998rq}
D.~J.~H. Chung, E.~W. Kolb, and A.~Riotto, ``{Production of massive particles during reheating},'' \href{https://dx.doi.org/10.1103/PhysRevD.60.063504}{{\em Phys. Rev. D} {\bfseries 60} (1999) 063504}, \href{https://arxiv.org/abs/hep-ph/9809453}{{\ttfamily arXiv:hep-ph/9809453}}.

\bibitem{Giudice:2000ex}
G.~F. Giudice, E.~W. Kolb, and A.~Riotto, ``{Largest temperature of the radiation era and its cosmological implications},'' \href{https://dx.doi.org/10.1103/PhysRevD.64.023508}{{\em Phys. Rev. D} {\bfseries 64} (2001) 023508}, \href{https://arxiv.org/abs/hep-ph/0005123}{{\ttfamily arXiv:hep-ph/0005123}}.

\bibitem{Garcia:2020eof}
M.~A.~G. Garcia, K.~Kaneta, Y.~Mambrini, and K.~A. Olive, ``{Reheating and Post-inflationary Production of Dark Matter},'' \href{https://dx.doi.org/10.1103/PhysRevD.101.123507}{{\em Phys. Rev. D} {\bfseries 101} no.~12, (2020) 123507}, \href{https://arxiv.org/abs/2004.08404}{{\ttfamily arXiv:2004.08404 [hep-ph]}}.

\bibitem{Garcia:2020wiy}
M.~A.~G. Garcia, K.~Kaneta, Y.~Mambrini, and K.~A. Olive, ``{Inflaton Oscillations and Post-Inflationary Reheating},'' \href{https://dx.doi.org/10.1088/1475-7516/2021/04/012}{{\em JCAP} {\bfseries 04} (2021) 012}, \href{https://arxiv.org/abs/2012.10756}{{\ttfamily arXiv:2012.10756 [hep-ph]}}.

\bibitem{Co:2020xaf}
R.~T. Co, E.~Gonzalez, and K.~Harigaya, ``{Increasing Temperature toward the Completion of Reheating},'' \href{https://dx.doi.org/10.1088/1475-7516/2020/11/038}{{\em JCAP} {\bfseries 11} (2020) 038}, \href{https://arxiv.org/abs/2007.04328}{{\ttfamily arXiv:2007.04328 [astro-ph.CO]}}.

\bibitem{Davidson:2000er}
S.~Davidson and S.~Sarkar, ``{Thermalization after inflation},'' \href{https://dx.doi.org/10.1088/1126-6708/2000/11/012}{{\em JHEP} {\bfseries 11} (2000) 012}, \href{https://arxiv.org/abs/hep-ph/0009078}{{\ttfamily arXiv:hep-ph/0009078}}.

\bibitem{Allahverdi:2002pu}
R.~Allahverdi and M.~Drees, ``{Thermalization after inflation and production of massive stable particles},'' \href{https://dx.doi.org/10.1103/PhysRevD.66.063513}{{\em Phys. Rev. D} {\bfseries 66} (2002) 063513}, \href{https://arxiv.org/abs/hep-ph/0205246}{{\ttfamily arXiv:hep-ph/0205246}}.

\bibitem{Drees:2022vvn}
M.~Drees and B.~Najjari, ``{Multi-species thermalization cascade of energetic particles in the early universe},'' \href{https://dx.doi.org/10.1088/1475-7516/2023/08/037}{{\em JCAP} {\bfseries 08} (2023) 037}, \href{https://arxiv.org/abs/2205.07741}{{\ttfamily arXiv:2205.07741 [hep-ph]}}.

\bibitem{Arnold:2002zm}
P.~B. Arnold, G.~D. Moore, and L.~G. Yaffe, ``{Effective kinetic theory for high temperature gauge theories},'' \href{https://dx.doi.org/10.1088/1126-6708/2003/01/030}{{\em JHEP} {\bfseries 01} (2003) 030}, \href{https://arxiv.org/abs/hep-ph/0209353}{{\ttfamily arXiv:hep-ph/0209353}}.

\bibitem{Migdal:1956tc}
A.~B. Migdal, ``{Bremsstrahlung and Pair Production at High Energies in Condensed Media},'' \href{https://dx.doi.org/10.1103/PhysRev.103.1811}{{\em Phys. Rev.} {\bfseries 103} (1956) 1811--1820}.

\bibitem{Baier:1996kr}
R.~Baier, Y.~L. Dokshitzer, A.~H. Mueller, S.~Peigne, and D.~Schiff, ``{Radiative energy loss of high-energy quarks and gluons in a finite volume quark - gluon plasma},'' \href{https://dx.doi.org/10.1016/S0550-3213(96)00553-6}{{\em Nucl. Phys. B} {\bfseries 483} (1997) 291--320}, \href{https://arxiv.org/abs/hep-ph/9607355}{{\ttfamily arXiv:hep-ph/9607355}}.

\bibitem{Baier:1996sk}
R.~Baier, Y.~L. Dokshitzer, A.~H. Mueller, S.~Peigne, and D.~Schiff, ``{Radiative energy loss and p(T) broadening of high-energy partons in nuclei},'' \href{https://dx.doi.org/10.1016/S0550-3213(96)00581-0}{{\em Nucl. Phys. B} {\bfseries 484} (1997) 265--282}, \href{https://arxiv.org/abs/hep-ph/9608322}{{\ttfamily arXiv:hep-ph/9608322}}.

\bibitem{Zakharov:1996fv}
B.~G. Zakharov, ``{Fully quantum treatment of the Landau-Pomeranchuk-Migdal effect in QED and QCD},'' \href{https://dx.doi.org/10.1134/1.567126}{{\em JETP Lett.} {\bfseries 63} (1996) 952--957}, \href{https://arxiv.org/abs/hep-ph/9607440}{{\ttfamily arXiv:hep-ph/9607440}}.

\bibitem{Arnold:2001ba}
P.~B. Arnold, G.~D. Moore, and L.~G. Yaffe, ``{Photon emission from ultrarelativistic plasmas},'' \href{https://dx.doi.org/10.1088/1126-6708/2001/11/057}{{\em JHEP} {\bfseries 11} (2001) 057}, \href{https://arxiv.org/abs/hep-ph/0109064}{{\ttfamily arXiv:hep-ph/0109064}}.

\bibitem{Arnold:2002ja}
P.~B. Arnold, G.~D. Moore, and L.~G. Yaffe, ``{Photon and gluon emission in relativistic plasmas},'' \href{https://dx.doi.org/10.1088/1126-6708/2002/06/030}{{\em JHEP} {\bfseries 06} (2002) 030}, \href{https://arxiv.org/abs/hep-ph/0204343}{{\ttfamily arXiv:hep-ph/0204343}}.

\bibitem{Arnold:2008zu}
P.~B. Arnold and C.~Dogan, ``{QCD Splitting/Joining Functions at Finite Temperature in the Deep LPM Regime},'' \href{https://dx.doi.org/10.1103/PhysRevD.78.065008}{{\em Phys. Rev. D} {\bfseries 78} (2008) 065008}, \href{https://arxiv.org/abs/0804.3359}{{\ttfamily arXiv:0804.3359 [hep-ph]}}.

\bibitem{Baier:2000sb}
R.~Baier, A.~H. Mueller, D.~Schiff, and D.~T. Son, ``{'Bottom up' thermalization in heavy ion collisions},'' \href{https://dx.doi.org/10.1016/S0370-2693(01)00191-5}{{\em Phys. Lett. B} {\bfseries 502} (2001) 51--58}, \href{https://arxiv.org/abs/hep-ph/0009237}{{\ttfamily arXiv:hep-ph/0009237}}.

\bibitem{Kurkela:2011ti}
A.~Kurkela and G.~D. Moore, ``{Thermalization in Weakly Coupled Nonabelian Plasmas},'' \href{https://dx.doi.org/10.1007/JHEP12(2011)044}{{\em JHEP} {\bfseries 12} (2011) 044}, \href{https://arxiv.org/abs/1107.5050}{{\ttfamily arXiv:1107.5050 [hep-ph]}}.

\bibitem{Kurkela:2014tea}
A.~Kurkela and E.~Lu, ``{Approach to Equilibrium in Weakly Coupled Non-Abelian Plasmas},'' \href{https://dx.doi.org/10.1103/PhysRevLett.113.182301}{{\em Phys. Rev. Lett.} {\bfseries 113} no.~18, (2014) 182301}, \href{https://arxiv.org/abs/1405.6318}{{\ttfamily arXiv:1405.6318 [hep-ph]}}.

\bibitem{Mrowczynski:2000ed}
S.~Mrowczynski and M.~H. Thoma, ``{Hard loop approach to anisotropic systems},'' \href{https://dx.doi.org/10.1103/PhysRevD.62.036011}{{\em Phys. Rev. D} {\bfseries 62} (2000) 036011}, \href{https://arxiv.org/abs/hep-ph/0001164}{{\ttfamily arXiv:hep-ph/0001164}}.

\bibitem{Mrowczynski:2004kv}
S.~Mrowczynski, A.~Rebhan, and M.~Strickland, ``{Hard loop effective action for anisotropic plasmas},'' \href{https://dx.doi.org/10.1103/PhysRevD.70.025004}{{\em Phys. Rev. D} {\bfseries 70} (2004) 025004}, \href{https://arxiv.org/abs/hep-ph/0403256}{{\ttfamily arXiv:hep-ph/0403256}}.

\bibitem{Arnold:2003rq}
P.~B. Arnold, J.~Lenaghan, and G.~D. Moore, ``{QCD plasma instabilities and bottom up thermalization},'' \href{https://dx.doi.org/10.1088/1126-6708/2003/08/002}{{\em JHEP} {\bfseries 08} (2003) 002}, \href{https://arxiv.org/abs/hep-ph/0307325}{{\ttfamily arXiv:hep-ph/0307325}}.

\bibitem{Romatschke:2003ms}
P.~Romatschke and M.~Strickland, ``{Collective modes of an anisotropic quark gluon plasma},'' \href{https://dx.doi.org/10.1103/PhysRevD.68.036004}{{\em Phys. Rev. D} {\bfseries 68} (2003) 036004}, \href{https://arxiv.org/abs/hep-ph/0304092}{{\ttfamily arXiv:hep-ph/0304092}}.

\bibitem{Rebhan:2004ur}
A.~Rebhan, P.~Romatschke, and M.~Strickland, ``{Hard-loop dynamics of non-Abelian plasma instabilities},'' \href{https://dx.doi.org/10.1103/PhysRevLett.94.102303}{{\em Phys. Rev. Lett.} {\bfseries 94} (2005) 102303}, \href{https://arxiv.org/abs/hep-ph/0412016}{{\ttfamily arXiv:hep-ph/0412016}}.

\bibitem{Braaten:1989mz}
E.~Braaten and R.~D. Pisarski, ``{Soft Amplitudes in Hot Gauge Theories: A General Analysis},'' \href{https://dx.doi.org/10.1016/0550-3213(90)90508-B}{{\em Nucl. Phys. B} {\bfseries 337} (1990) 569--634}.

\bibitem{Frenkel:1989br}
J.~Frenkel and J.~C. Taylor, ``{High Temperature Limit of Thermal QCD},'' \href{https://dx.doi.org/10.1016/0550-3213(90)90661-V}{{\em Nucl. Phys. B} {\bfseries 334} (1990) 199--216}.

\bibitem{Blaizot:2001nr}
J.-P. Blaizot and E.~Iancu, ``{The Quark gluon plasma: Collective dynamics and hard thermal loops},'' \href{https://dx.doi.org/10.1016/S0370-1573(01)00061-8}{{\em Phys. Rept.} {\bfseries 359} (2002) 355--528}, \href{https://arxiv.org/abs/hep-ph/0101103}{{\ttfamily arXiv:hep-ph/0101103}}.

\bibitem{Blaizot:2015jea}
J.-P. Blaizot and Y.~Mehtar-Tani, ``{Energy flow along the medium-induced parton cascade},'' \href{https://dx.doi.org/10.1016/j.aop.2016.01.002}{{\em Annals Phys.} {\bfseries 368} (2016) 148--176}, \href{https://arxiv.org/abs/1501.03443}{{\ttfamily arXiv:1501.03443 [hep-ph]}}.

\bibitem{Blas:2026yqb}
D.~Blas, S.~Gasparotto, H.~Murayama, J.~Sch{\"u}tte-Engel, and J.~Zosso, ``{Low-Frequency Gravitational Bremsstrahlung and Memory in a Medium},'' \href{https://arxiv.org/abs/2608.24994}{{\ttfamily arXiv:2608.24994 [hep-ph]}}.

\bibitem{AbraaoYork:2014hbk}
M.~C. Abraao~York, A.~Kurkela, E.~Lu, and G.~D. Moore, ``{UV cascade in classical Yang-Mills theory via kinetic theory},'' \href{https://dx.doi.org/10.1103/PhysRevD.89.074036}{{\em Phys. Rev. D} {\bfseries 89} no.~7, (2014) 074036}, \href{https://arxiv.org/abs/1401.3751}{{\ttfamily arXiv:1401.3751 [hep-ph]}}.

\bibitem{Kurkela:2018oqw}
A.~Kurkela and A.~Mazeliauskas, ``{Chemical equilibration in weakly coupled QCD},'' \href{https://dx.doi.org/10.1103/PhysRevD.99.054018}{{\em Phys. Rev. D} {\bfseries 99} no.~5, (2019) 054018}, \href{https://arxiv.org/abs/1811.03068}{{\ttfamily arXiv:1811.03068 [hep-ph]}}.

\bibitem{Boguslavski:2024kbd}
K.~Boguslavski and F.~Lindenbauer, ``{Soft-gluon exchange matters: Isotropic screening in QCD kinetic theory},'' \href{https://dx.doi.org/10.1103/PhysRevD.110.074017}{{\em Phys. Rev. D} {\bfseries 110} no.~7, (2024) 074017}, \href{https://arxiv.org/abs/2407.09605}{{\ttfamily arXiv:2407.09605 [hep-ph]}}.

\bibitem{Iancu:2015bip}
E.~Iancu and B.~Wu, ``{Thermalization of mini-jets in a quark{\textendash}gluon plasma},'' \href{https://dx.doi.org/10.1016/j.nuclphysa.2016.04.028}{{\em Nucl. Phys. A} {\bfseries 956} (2016) 581--584}, \href{https://arxiv.org/abs/1512.09353}{{\ttfamily arXiv:1512.09353 [hep-ph]}}.

\end{thebibliography}\endgroup

\end{document}